\documentclass[a4paper,twocolumn,11pt,accepted=2024-09-18]{quantumarticle}
\pdfoutput=1
\usepackage{ltxgrid}
\usepackage[numbers,sort&compress]{natbib}
\usepackage{amsmath}
\usepackage{amssymb}

\usepackage[pdftex, pdftitle={Article}, pdfauthor={Author}]{hyperref} 
\usepackage{subfigure}
\usepackage{float}

\begin{document}

\title{Efficient Parameter Optimisation for Quantum Kernel Alignment: A Sub-sampling Approach in Variational Training}

\author{M. Emre Sahin}
\affiliation{The Hartree Centre, STFC, Sci-Tech Daresbury, Warrington, WA4 4AD, United Kingdom}
\author{Benjamin C. B. Symons}
\affiliation{The Hartree Centre, STFC, Sci-Tech Daresbury, Warrington, WA4 4AD, United Kingdom}
\author{Pushpak Pati}
\affiliation{IBM Research -- Zurich, S\"aumerstrasse 4, R\"uschlikon, CH–8803, Switzerland}

\author{Fayyaz Minhas}
\affiliation{Tissue Image Analytics Centre, Warwick Cancer Research Centre, Department of Computer Science, University of Warwick, Coventry, CV4 7AL, United Kingdom}

\author{Declan Millar}
\affiliation{IBM Research Europe - UK, Hursley, Winchester, SO21 2JN, United Kingdom}

\author{Maria Gabrani}
\affiliation{IBM Research -- Zurich, S\"aumerstrasse 4, R\"uschlikon, CH–8803, Switzerland}

\author{Stefano Mensa}
\email{stefano.mensa@stfc.ac.uk}
\affiliation{The Hartree Centre, STFC, Sci-Tech Daresbury, Warrington, WA4 4AD, United Kingdom}

\author{Jan Lukas Robertus}
\email{j.robertus@imperial.ac.uk}
\affiliation{Royal Brompton and Harefield Hosptials, Department of Histopathology, Guy’s and St Thomas’ NHS Foundation Trust, Sydney Street, London SW3 6NP United Kingdom}
\affiliation{National Heart and Lung Institute, Imperial College London, Guy Scadding Building, Dovehouse St, London SW3 6LY United Kingdom}
\thanks{SM and JLR share joint last authorship and contributed equally to this work. For all communications please address both authors.}


\begin{abstract}

Quantum machine learning with quantum kernels for classification problems is a growing area of research. Recently, quantum kernel alignment techniques that parameterise the kernel have been developed, allowing the kernel to be trained and therefore aligned with a specific dataset. While quantum kernel alignment is a promising technique, it has been hampered by considerable training costs because the full kernel matrix must be constructed at every training iteration. Addressing this challenge, we introduce a novel method that seeks to balance efficiency and performance. We present a sub-sampling training approach that uses a subset of the kernel matrix at each training step, thereby reducing the overall computational cost of the training. In this work, we apply the sub-sampling method to synthetic datasets and a real-world breast cancer dataset and demonstrate considerable reductions in the number of circuits required to train the quantum kernel while maintaining classification accuracy.

\end{abstract}

\maketitle

\section{Introduction}\label{sec1}

Kernel methods are a fundamental concept in Machine Learning (ML), with a rich history that spans several decades, revolutionising the field by enabling the effective handling of complex, nonlinear relationships in data~\cite{hofmann2008kernel}. The introduction of the ``kernel trick" in the 1960s laid the foundation~\cite{aiserman1964theoretical}, leading to the development of Support Vector Machines (SVMs) in the 1990s~\cite{boser1992training}, which popularized the concept. Over time, a diverse range of kernel functions have emerged, broadening their utility across many domains. Another milestone was the introduction of the concept of kernel alignment \cite{cristianini2001kernel}. This method adjusts a kernel function in relation to specific data, thereby limiting the generalisation error~\cite{wang2015overview, cortes2012algorithms}. While facing challenges related to scalability~\cite{dai2014scalable}, kernel methods continue to be relevant, coexisting with deep learning in solving complex problems.

The field of ML stands to benefit from quantum algorithms, offering the potential for improved scalability and accuracy in solving complex computational tasks~\cite{liu2021rigorous, schuld2021machine, biamonte2017quantum, riste2017demonstration}. A particularly promising approach in this field is the utilization of quantum fidelity kernels, a quantum-enhanced version of kernel methods~\cite{schuld2019quantum,mengoni2019kernel, havlivcek2019supervised, liu2021rigorous, huang2021power}. These kernels quantify the overlap between quantum states and leverage the high dimensionality of quantum systems to learn and capture complex patterns in data. A few relevant practical applications of quantum machine learning with kernel methods can be seen in Refs.~\cite{mensa2023quantum,wu2021application,wu2023quantum,gentinetta2022complexity,gentinetta2023quantum, miyabe2023quantum}.

Despite the promise of quantum kernel methods, there is scepticism about their practical application due to challenges associated with scalability. These challenges stem from the convergence of quantum kernel values which depends on factors such as data embedding expressibility, the number of qubits, global measurements, entanglement, and noise, which often demand increased measurements for effective training \cite{thanasilp2022exponential, cerezo2022challenges}. In response to these challenges, several techniques have been introduced \cite{huang2021power, kubler2021inductive, shirai2021quantum}, which, along with bandwidth adjustments \cite{canatar2022bandwidth, shaydulin2022importance}, aim to enhance generalization in quantum kernels, particularly at higher qubit counts.

Quantum kernel alignment (QKA) adapts classical kernel alignment to utilize quantum kernels, enhancing the model's precision by aligning more effectively with core data patterns, as outlined in studies by \cite{hubregtsen2022training, glick2021covariant, gentinetta2023quantum}. Despite this advancement, the scaling in terms of the number of circuits required to align quantum kernels with QKA presents a considerable challenge. In the noiseless case the number of circuits at each training step scales quadratically with dataset size and in the presence of noise it has been shown to scale quartically \cite{gentinetta2022complexity}, which can quickly become prohibitive for larger datasets \cite{miyabe2023quantum, thanasilp2022exponential}. Addressing this complexity with respect to dataset size is crucial for enabling the scalability of QKA methods. One way to address the computational challenges of QKA is to adapt the PEGASOS algorithm \cite{gentinetta2023quantum}. In reference \cite{gentinetta2023quantum}, they demonstrate that the complexity of this method for an $\epsilon$-accurate classifier with a dataset size of $m$ is $O(min\{ m^2/\epsilon^6, 1/ \epsilon^{10}\})$ as compared to $O(m^{4.67} / \epsilon^2)$ for standard quantum kernel alignment in the presence of finite sampling noise. Indeed the PEGASOS-based algorithm scales better than the standard approach with respect to $m$ which makes it attractive for larger datasets but, it comes at the cost of worse scaling with respect to $\epsilon$.

In this work, we aim to improve the scaling with respect to the dataset size of QKA methods by employing sub-sampling techniques during training. Specifically, we utilize a subset of the kernel matrix, termed the sub-kernel, in each training step. This approach aims to reduce the computational burden associated with training, while still effectively learning kernel parameters. As such, the prospective use of our sub-sampling approach opens up new avenues for scalable and efficient quantum machine learning algorithms.

 \begin{figure*}[t!]
  \centering
  \includegraphics[width=0.6\textheight]{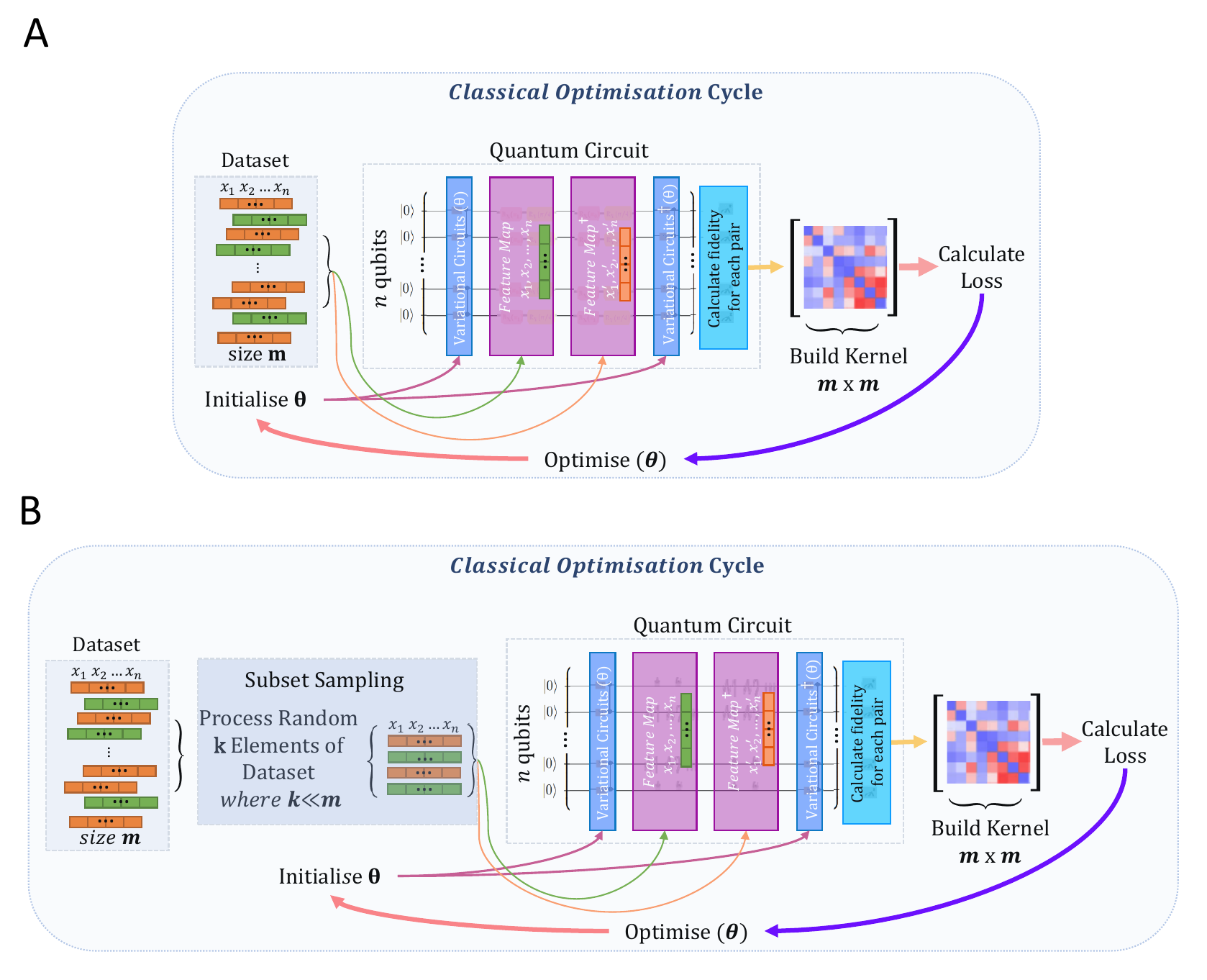}
  \caption{ \textbf{A)} Quantum kernel alignment uses \textit{n} qubits to encode the dataset into a feature map, estimate the kernel, and optimize parameters $\boldsymbol{\theta}$ until kernel separation improves. \textbf{B)} In the sub-sampling method, subsets with $k$ elements of the dataset are similarly processed using \textit{n} qubits. This cycle repeats with different subsets, optimizing $\boldsymbol{\theta}$ until the entire dataset is sampled, yielding an optimized kernel. This will end with the estimation of the full dataset, optimised kernel.}
  \label{fig:figure1}
\end{figure*}

\section{Methodology}\label{methodology}

Classical kernel methods compute elements of a kernel matrix $K_{x,x'}$ using a kernel function $K(\boldsymbol{x},\boldsymbol{x}')$ that measures the distance between two data points $\boldsymbol{x}$, $\boldsymbol{x}' \in X$. For certain choices of distance measure it is possible to use the kernel trick to express the kernel function as an inner product in some vector space $V$. If the kernel function can be written as a feature map $f:X \rightarrow V$, then $V$ is known as the feature space and the kernel function can be written as an inner product on $V$, $K(\boldsymbol{x},\boldsymbol{x}') = \langle f(\boldsymbol{x}), f(\boldsymbol{x}')\rangle_V$. We can see from this definition that the kernel function is a measure of similarity between data points that takes on values $[ 0, 1 ]$ e.g. if $x=x'$ then the inner product is one, whereas if the maps $f(x)$, $f(x')$ result in orthogonal vectors then the inner product is zero.

Quantum kernel methods can be defined analogously to classical kernel methods. A quantum feature map, \( \phi(\boldsymbol{x}) \), is used to transform classical data \( \boldsymbol{x} \) into a corresponding quantum state \( |\phi(\boldsymbol{x})\rangle \). The fidelity kernel between two classical data points, \( \boldsymbol{x} \) and \( \boldsymbol{x}' \), is computed using the overlap of the related quantum states $K(\boldsymbol{x}, \boldsymbol{x}') = | \langle\psi ( \boldsymbol{x})|\psi ( \boldsymbol{x}')\rangle |^2$, where \( |\psi (\boldsymbol{x})\rangle = U_{\phi}(\boldsymbol{x}) |0\rangle^{\otimes n} \) and $U_{\phi}(\boldsymbol{x})$ is the unitary that implements the quantum feature map. The quality of such kernels can be further improved by quantum kernel alignment techniques that parameterise the kernel and use a variational approach to optimize for the kernel for a specific dataset as suggested in Refs. \cite{hubregtsen2022training, glick2021covariant} (see Fig.~\ref{fig:figure1}). QKA adds a variational layer $U(\boldsymbol{\theta})$ to the kernel, where \( \boldsymbol{\theta} \) are trainable parameters. The elements of the fidelity kernel matrix are computed in the same way, but now depend on $\boldsymbol{\theta}$ as shown in Equation \ref{kernel-align},

  \begin{equation}
  \label{kernel-align}
    K(\boldsymbol{x}, \boldsymbol{x}', \boldsymbol{\theta}) = 
    |\langle\psi (\boldsymbol{\theta}, \boldsymbol{x})|\psi (\boldsymbol{\theta}, \boldsymbol{x}')\rangle |^2,
  \end{equation}

where $|\psi (\boldsymbol{\theta}, \boldsymbol{x})\rangle = U_{\phi}(\boldsymbol{x}) U(\boldsymbol{\theta}) |0\rangle^{\otimes n}$. Note that, while we focus on kernels of the form shown in Equation \eqref{kernel-align}, in principle our method generalises to other choices of kernel. 

Similar to classical kernel alignment \cite{cristianini2001kernel}, the training of the variational quantum kernel can be seen as finding a kernel that minimizes the upper bound of the SVM generalization error. This is equivalent to minimising a loss function $L$,

\begin{equation}
L(\boldsymbol{\theta},a) = \sum_{i} a_i - 0.5 \sum_{i,j} a_i a_j y_{i} y_{j} K(\boldsymbol{\theta},x_i, x_j),
\label{hinge-loss}
\end{equation}

where $a_i$ and $a_j$ are the optimal Lagrange multipliers, and $y$ are the labels of the binary classification problem. The hinge loss, as a function of the variational kernel \( K(\boldsymbol{\theta}) \), is not generally convex, hence the loss landscape may include local minima, complicating the training process. 

While kernel alignment has been shown to improve the performance of kernel methods, it comes at a cost: for every iteration of the training process, the full kernel must be constructed which for a dataset with $m$ points requires $m^2$ inner product evaluations. For quantum kernel alignment this translates to $m^2$ circuits per training iteration which, as the dataset grows, can become prohibitively costly. In this paper, we present a novel method that utilises a so-called sub-sampling approach to speed-up the model training process by reducing the number of circuits per training iteration.

The full kernel, $K_{f}(\boldsymbol{\theta})$ is defined on the full dataset $D$ whereas the sub-kernel $K_{s}(\boldsymbol{\theta})$ is defined using a subset of the data $D_s \subset D$ that is randomly sampled from $D$. The sub-kernel is used during the variational optimisation phase in order to find optimal parameters $\boldsymbol{\theta}_{opt}$. Once obtained, the parameters $\boldsymbol{\theta}_{opt}$ are used to construct the full fidelity kernel $K_{f}(\boldsymbol{\theta}_{opt})$, which is then used by a Support Vector Classifier (SVC). The benefit of this approach is that fewer circuits are required to build a sub-kernel than the full kernel, therefore training using a sub-kernel can speed-up the training process by a factor that is at worst constant and at best removes the dependence of scaling on $m$. The method, schematically represented in Fig.~\ref{fig:figure1} B, can be summarised as follows:

\begin{itemize}
\item \textbf{Variational parameter initialisation:} the variational parameters $\boldsymbol{\theta}$ of the quantum circuit are initialised as $\boldsymbol{\theta}_{init}$, giving an initial state $|\psi(\boldsymbol{\theta}_{init})\rangle$.

\item \textbf{Variational parameters and loss function optimisation loop:} execute the optimisation loop and for each iteration, execute the following sub-steps:

    \begin{itemize}
    \item \textbf{Dataset subsampling:} initiate by randomly selecting a distinct subset of \( k \) data points from the full dataset and employ Equation (1) to construct a sub kernel, \( K_s \). Note that each time a new subset is selected, it is composed of previously unselected points until the entire dataset is exhausted, at which point previously selected points may be used again.    

    \item \textbf{Loss function estimation:} compute the loss function $L$ using $K_s$ (Equation \ref{hinge-loss}).
    
    \item \textbf{Loss function optimisation:} perform an optimisation step on $L$ to update the parameters $\boldsymbol{\theta}$.
    \end{itemize}

\item \textbf{Full kernel estimation:} utilise the optimised parameters $\boldsymbol{\theta}_{opt}$ to compute the full quantum kernel $K_f(\boldsymbol{\theta}_{opt})$.

\item \textbf{Training of classifier:} implement the full kernel in an SVC model for binary classification.

\end{itemize}

When introducing sub-sampling into the optimisation process, instead of computing the loss \(L_f = L(D)\) over the full dataset \(D\), the loss is evaluated on a sub-sample of size $k$, \(L_s = L(D_s)\). This introduces a sampling error \(\epsilon_s\). Assuming the sub-samples are i.i.d., the expected sampling error can be bounded by \(\mathbb{E}[|\epsilon_s|] \leq O\left(\frac{\sigma_k}{\sqrt{s}}\right)\), where $\sigma_k^2$ is the variance of the distribution of losses for a sub-sample of size $k$ and $s$ is the number of samples. In the limit that the sub-sample size is equal to the full dataset, then the variance in the loss goes to zero. We cannot say anything general about $\sigma_k$ as it will depend on the specifics of the dataset and the choice of $k$ among other things. However, we expect that sub-sampling will introduce some non-zero variance that could be large. The error in the loss can be controlled by taking multiple sub-samples and averaging them in each step of the training. Nonetheless, the variance in the loss will likely result in more training iterations being required to reach some fixed convergence criterion, compared to the full kernel training method.\\
\indent In the case of training with a full kernel, every iteration involves processing all $m$ data points. The full kernel matrix contains $m^2$ entries which correspond to $m^2$ evaluations of the kernel function and therefore, assuming no noise, a query complexity of $O(m^2)$ per training iteration. In the presence of finite sampling noise, it has been shown that sampling overhead results in a considerably worse complexity of $O(m^{4.67}/\epsilon^2)$ per training iteration for a kernel that is $\epsilon$-close to the infinite shot kernel \cite{gentinetta2022complexity}. For the remainder of this work, we consider the noise-free case but note that the speed-ups presented may in fact improve in the presence of noise.\\
\indent Denoting the number of iterations required for convergence as $T$, the overall complexity for the full kernel training is $O(m^2 T)$. On the other hand, with sub-kernel training, we only process a subset of the data points of size $k$ at each iteration. Given $T'$ training iterations and $s$ sub-kernels per iteration, the overall complexity of the sub-sampling approach is $O(s k^2 T')$. As discussed previously, it is likely that $T' > T$ meaning it is not guaranteed that the complexity is reduced. It is evident that the upper bound on $T'$ is given by $T' < \frac{m^2}{sk^2} T$. Note that, while increasing $s$ decreases the upper bound on $T'$, it also decreases the loss error $\epsilon_s$ in each iteration which in turn is likely to decrease the actual value of $T'$. Assuming that $k << m$ and that $s < m^2/k^2$, then $T'$ is allowed to be a potentially large multiple of $T$. This opens up the possibility for considerable reductions in run time when training using the sub-sampling approach. Note that, if $s$ and $k$ are chosen to be constants independent of $m$ and $T'$ does not scale with $m$, then the overall complexity of the training process is independent of $m$. While there are no \textit{a priori} guarantees that $T'$ will satisfy the bound given above, our results shown in the next section demonstrate significant speed-ups with little to no loss of classification accuracy relative to the full kernel method for choices of $s$ and $k$ that are independent of $m$.

\section{Results from Numerical Simulations and Discussion}\label{Results from numerical simulations}

In this section we demonstrate our novel method by applying it to several test cases, showcasing the reductions in optimised kernel estimation times as well as retained - or improved - classification accuracy. We start by describing the setup of our tests and then we present results using two synthetic and one real-world dataset as examples.\\
\indent Our experiments involved two variational ansatze: Real Amplitudes (RA) and Hardware Efficient (HE) \cite{Farhi2014}, often referred to as Efficient SU2 ansatze. We used three different optimisers including Gradient Descent (GD) \cite{Bottou2010}, SPSA (Simultaneous Perturbation Stochastic Approximation) \cite{Spall1998}, ADAM (Adaptive Moment Estimation) \cite{Kingma2014}. We employed multiple commonly used ansatze and optimisers to test whether or not these choices impact the performance of the sub-sample method. We used different feature maps to encode the data according to the dataset and specify each in the relevant sections. To ensure fairness and reproducibility, the same initial variational parameters are used for every run.
 Unless stated otherwise, the results presented were obtained from statevector simulations using Qiskit \cite{javadi2024quantum}. The standard deviation of the outcomes through 10-fold cross-validation across all simulations was calculated. We maintained the same test and train subsets for all SVC methods to facilitate an accurate direct comparison between the different algorithms. All experiments were conducted five times, and the results shown are the averages over the values obtained.\\
\indent In the results presented here, we use the number of queries as a figure of merit. This has the advantage of being a platform-independent metric. Here, a query refers to a single execution of a fidelity calculation i.e., a single circuit. By comparing the number of queries required to reach the optimal solution for each method, we can determine the relative speed-up one method has over another. Specifically, we define speed-up to be the ratio of the number of queries required for a full kernel compared to a given sub-kernel of size $k$.

\subsection{Synthetic Datasets}

\subsubsection{Benchmarking for Second-order Pauli-Z evolution circuit}

For the empirical evaluation of our method, we use a synthetic dataset that is constructed in line with the principles set out by Havlíček et al. \cite{havlivcek2019supervised}, where the dataset is specifically designed to be fully separable through a second-order Pauli-Z evolution. The feature map transforms uniformly distributed vectors into quantum states, and labels are attributed based on the inner products of these quantum states. A separation gap of 0.2 is used to set the labeling threshold, ensuring a balanced yet challenging classification problem. The dataset comprises 96 training data points and 32 test data points, each represented by a 2-qubit quantum state. 

\begin{figure*}[t]
  \centering
  \includegraphics[width=0.8\textwidth]{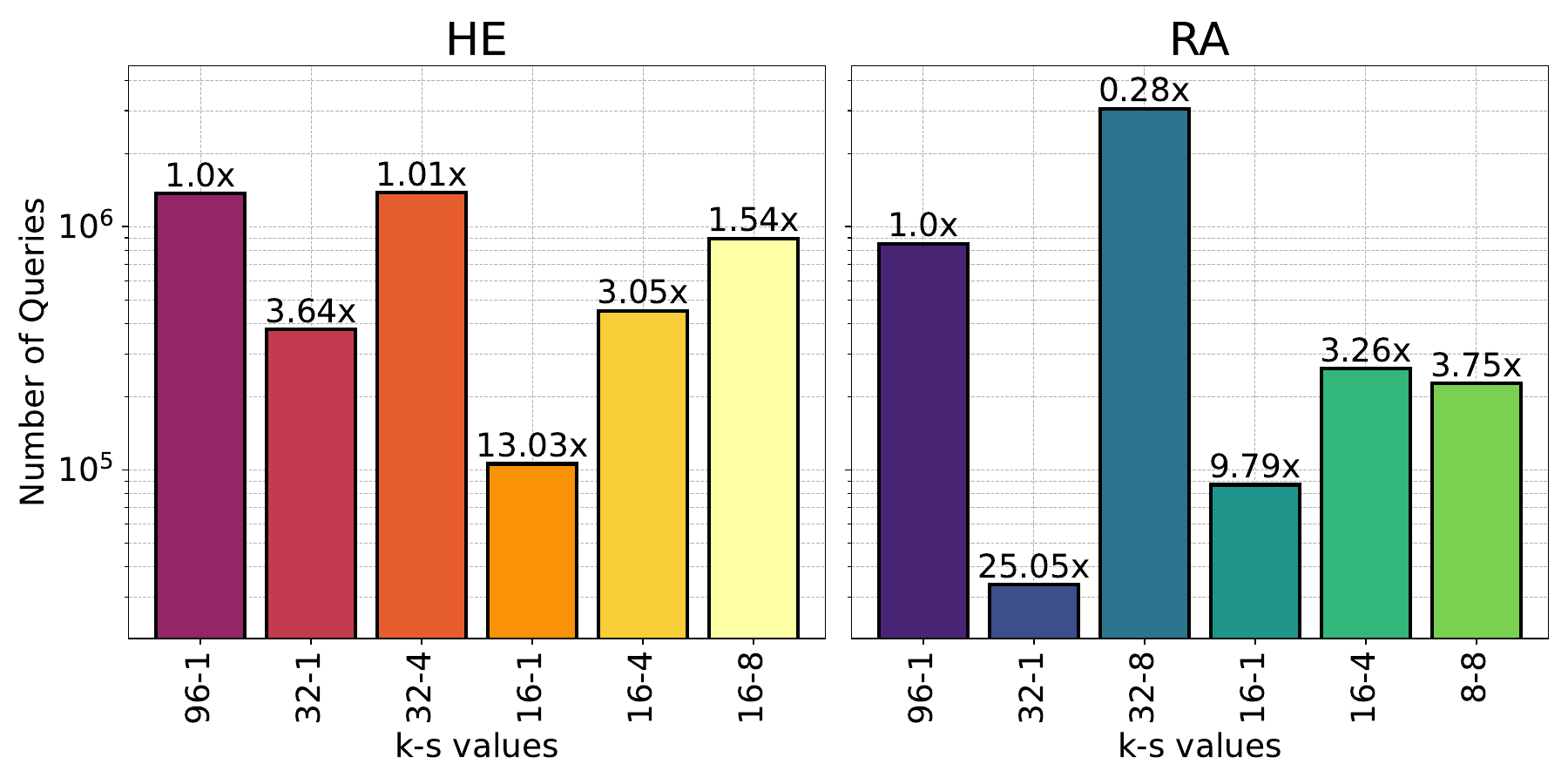}
  \includegraphics[width=0.8\textwidth]{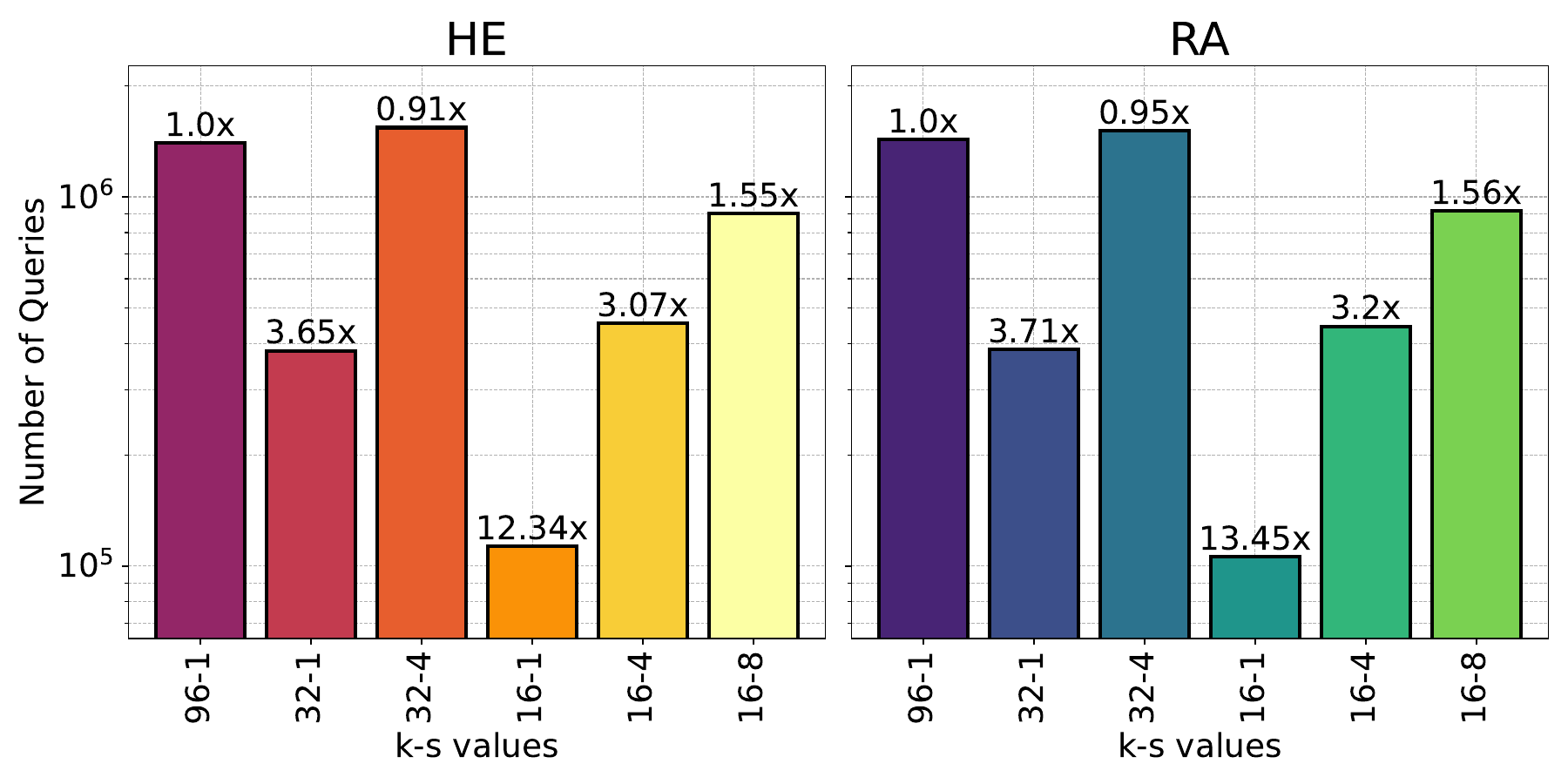}
  \caption{Total number of queries and speed-up relative to full kernel for the top 12 results using the SPSA optimizer, grouped by Hardware Efficient (HE) ansatze and Real Amplitudes (RA) ansatze. \textbf{Upper panels:} the range of ROC AUC values is 0.975-0.990 and the F1 scores range from 0.856-0.931 on the ZZ dataset (see Table \ref{apx-spsa-zz} for full results). \textbf{Lower panels:} all ROC AUC values are 1.0 and the F1 values range from 0.969-1.0 on the Coset dataset (see Table \ref{apx-spsa-coset} for full results). In both panels, $k$ is the sub-kernel size, and $s$ is the number of samples. Results are ordered by descending $k$.}
  \label{fig:merged-results}
\end{figure*}

In Fig. \ref{fig:merged-results}, we show the results obtained from statevector simulation using the SPSA optimiser with different ansatze, sub-samples sizes and the corresponding number of queries and relative speed-up compared to the full kernel evaluation. In this case, we set the maximum number of training iterations to 200 and if the optimiser reaches this point then we pick the point with the lowest loss. If the point with the lowest loss occurs before the final point, we consider this to be the stopping point of the run. In practice, a different stopping criterion such as a loss threshold could be chosen to avoid running more iterations than necessary. All 3 optimisers tested performed comparably, for the full list of results, see the Tables in Appendix B.

For the HE Ansatz results shown in the left panel of Fig. \ref{fig:merged-results}, the full kernel achieves an ROC AUC of 0.98 and an F1 score of 0.93. For all choices of $k$ and $s$, the ROC AUC is slightly improved relative to the full kernel, with an average improvement of 0.7\%. The F1 score achieved is identical to the full kernel in all cases apart from $k=16$, $s=1$ which sees an 8.2\% drop in F1 score. For the RA Ansatz results shown in the right panel of Fig. \ref{fig:merged-results}, the full kernel achieves the same ROC AUC and F1 score as for the HE Ansatz. The parameters $k=8$ and $s=8$ are the only choice for which the ROC AUC and F1 decrease relative to the full kernel (by 0.51\% and 3.5\% respectively). For all other parameter choices shown the F1 score is identical to the full kernel score and the ROC AUC increases by 0.51-1.02\%.\\
\indent The results demonstrate that the sub-sampling method can achieve similar performance using fewer circuits. The number of queries shown in Figure \ref{fig:merged-results} is the total number of queries, i.e. the number of queries per iteration multiplied by the number of iterations. We therefore see that, despite the sub-sampling method requiring more iterations to converge the loss function compared to the full kernel method (see Appendix \ref{Apx-loss} Figs \ref{fig:Synthetic-SPSA-LOSS} - \ref{fig:COSETS-GD-LOSS}), the total number of queries is still reduced considerably in almost all cases, resulting in up to x25 speed-up (reduction in number of queries). The only case for which there is a slowdown is for the RA ansatz when $k=32$ and $s=8$. This is perhaps unsurprising given that $m=96$ and so with these parameters the ratio $m^2/sk^2$ is relatively close to 1. Furthermore, the accuracy of the classification is not deteriorated by the sub-sampling method, showing comparable or even improved accuracy compared to the calculation of the full kernel ($k=96$ and $s=1$ for both HE and RA ansatze).\\
\indent We also performed a simple hardware experiment for this test case with the SPSA optimiser that can be seen in Appendix \ref{Apx-hardware} Table \ref{table:hw}. The experiments were performed on IBM Quantum hardware, Nairobi, using 2 qubits with 100 shots and no error mitigation. While we cannot draw general conclusions from these experiments, it is interesting to note that the speed-ups are realised even when measuring run time rather than queries. Furthermore, while the performance in terms of ROC AUC and F1 is lower in all cases (full and sub-kernel), we observe several cases where the sub-sampling method improves classification accuracy relative to the full kernel. This is some initial evidence that the method works on real hardware in the presence of noise.

\subsubsection{Learning Problem Labeling Cosets with Error (LCE)}

The second synthetic dataset we tested was the learning problem Labeling Cosets with Error (LCE) dataset, which was crafted to evaluate the accuracy of QML algorithms. The dataset was generated with a 27-qubit superconducting quantum processor \cite{glick2021covariant} which is split by 96 training points and 32 test points, each a 7-qubit quantum state defined by Euler angles. These angles specify elements of two separate cosets, forming the basis for binary classification. We apply the same feature map as used in  \cite{glick2021covariant}.

Similarly to Section 3.1.1, we present the results obtained using the SPSA optimiser in Figure \ref{fig:merged-results} (see Appendix \ref{Apx-Num} Tables for the full set of results). Also in this case, we observe over a $\times10$ speed-up, with the number of queries reduced as we reduce $k$. For both the HE and RA ansatzes in Fig. \ref{fig:merged-results}, the full kernel ROC AUC and F1 score are 1. For all choices of $k$ and $s$ the ROC AUC remains at 1. All F1 scores are also 1 apart from the choice of $k=16$, $s=8$ which sees a small reduction to 0.969. We therefore see that the classification accuracy of the method is, once again, not affected by the method and agrees with our previous discussion in Section 3.1.1.

\subsection{Real-World Dataset Experiment: Computational Pathology Breast Cancer Dataset}

\begin{figure*}[t]
  \centering
  \includegraphics[width=1\textwidth]{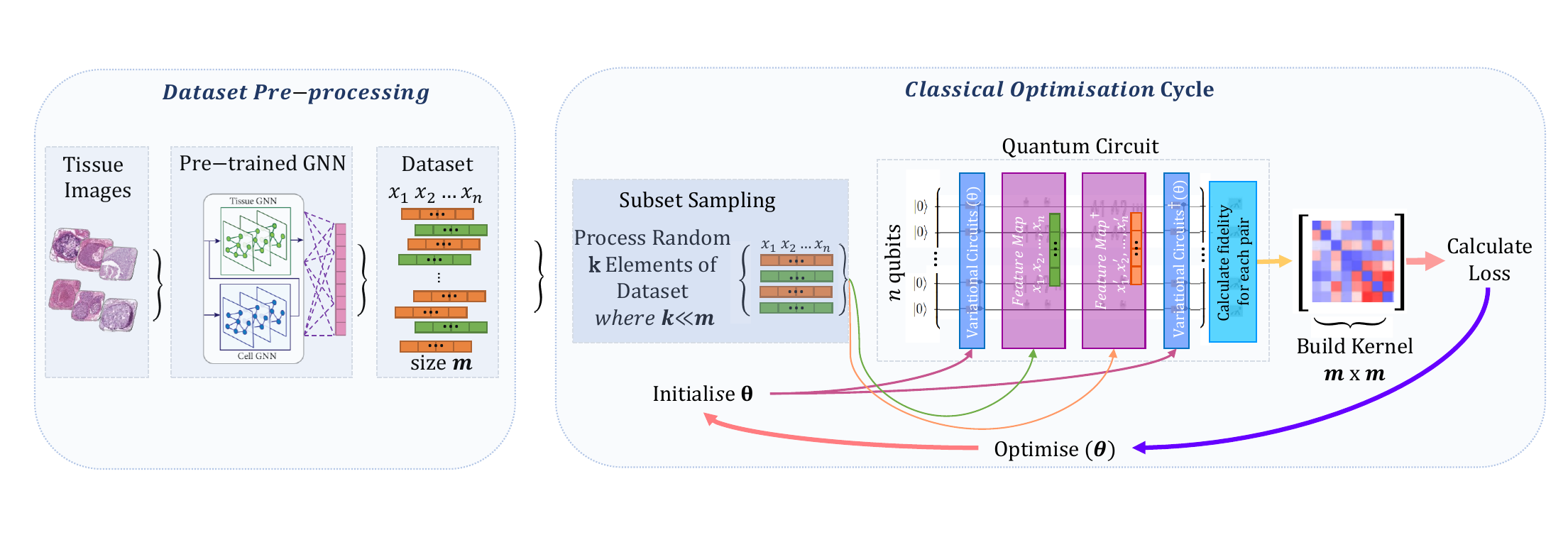}
  \caption{Combined GNN-extracted \cite{pati2022hierarchical} hierarchical cellular features with our sub-sampling method to optimise the creation of a quantum kernel used to classify subtypes of cancer. The GNN-extracted embeddings are used to create the dataset that will be sampled using the methodology discussed in this work. The output of the workflow is a fully variationally optimised quantum kernel.}
  \label{fig:figure2}
\end{figure*}

In the previous two subsections, we highlighted the strengths of our method with two synthetic datasets, showing that it's possible to train a quantum kernel faster, and without compromising accuracy using a sub-sampling approach. To further substantiate these benefits, we tested our method on a real-world application where high classification accuracy is non-negotiable - computational pathology for detecting and subtyping cancer. We chose this as an example dataset because it is a classification problem that is known to be classically challenging and where small improvements in accuracy can have a significant real-world impact. While the aim of this work is not to improve accuracy over the state-of-the-art classical methods, we believe the speed-ups presented in this work may help to facilitate future progress towards this goal.

Accurate cancer detection and subtyping (i.e. the process of categorising a cancer into a more specific group based on its characteristics) \cite{o2022precision}, involves analyzing complex data from diagnostic images using Graph Neural Networks (GNNs). Typically the GNN is used to extract features that are then passed to a classifier \cite{ahmedt2022survey} (see Fig. \ref{fig:figure2}). These classifiers have been shown to have high accuracy in classifying invasive vs non-invasive cancers but lower accuracy when classifying subtypes \cite{pati2022hierarchical}.

In this work, we take the features extracted from a GNN using the framework as described in \cite{pati2022hierarchical} on breast cancer (BRACS) data. In the dataset, there are three distinct pathological benign lesions including Usual Ductal Hyperplasia (UDH), Flat Epithelial Atypia (FEA) and Atypical Ductal Hyperplasia (ADH), and malignant lesions including Ductal Carcinoma In Situ (DCIS) and Invasive Carcinoma (IC) besides Normal breast tissue.
The subtypes are divided into two complex categories. The first category compares ADH and FEA  to DCIS (1566 training points, 275 test points). In the second category  Normal, and UDH are compared to ADH, FEA and DCIS (2797 Training points, 524 Test points).

In our tests, we used a 10-qubit feature map inspired by an instantaneous quantum polynomial (IQP) kernel \cite{havlivcek2019supervised, huang2021power} with an optimal bandwidth parameter $c_{opt} \approx 2/n$ \cite{shaydulin2022importance,canatar2022bandwidth} to reduce effects of exponential concentration quantum kernels \cite{thanasilp2022exponential}.

\begin{equation}
\begin{split}
&U_{IQP}(x) = U_Z(x) \otimes H^n U_Z(x) \otimes H^n,\\
&U_Z(x) = \exp \left( c \sum_{j=1}^{n} x_j Z_j + c^2 \sum_{j,j'=1}^{n} x_j x_{j'} Z_j Z_{j'} \right),
\end{split}
\end{equation}

where c is the bandwidth, \( H \) and \( Z \) denote the Hadamard and Pauli Z gates respectively.

\begin{widetext}
\begin{figure}[t]
\begingroup\renewcommand{\caption}[2]{}
    \centering
\resizebox{\textwidth}{!}{
    \centering
    \begin{minipage}[t]{0.2\textwidth}
        \centering
        \includegraphics[width=\textwidth]{Rebuttal-for-quantum/plots/pathology/legend.pdf}
    \end{minipage}}
        
\resizebox{\textwidth}{!}{    
    \begin{minipage}[t]{0.05\textwidth}
        \centering
        \includegraphics[width=\textwidth]{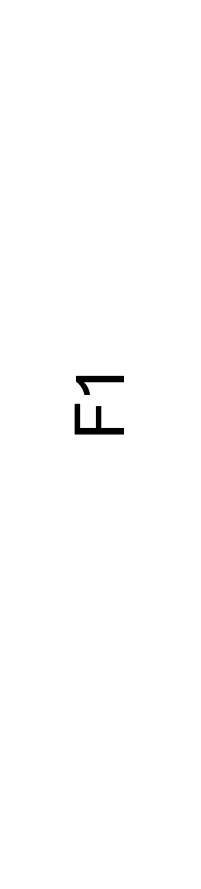}
    \end{minipage}
    \begin{minipage}[t]{0.4\textwidth}
        \centering
        \includegraphics[width=\textwidth]{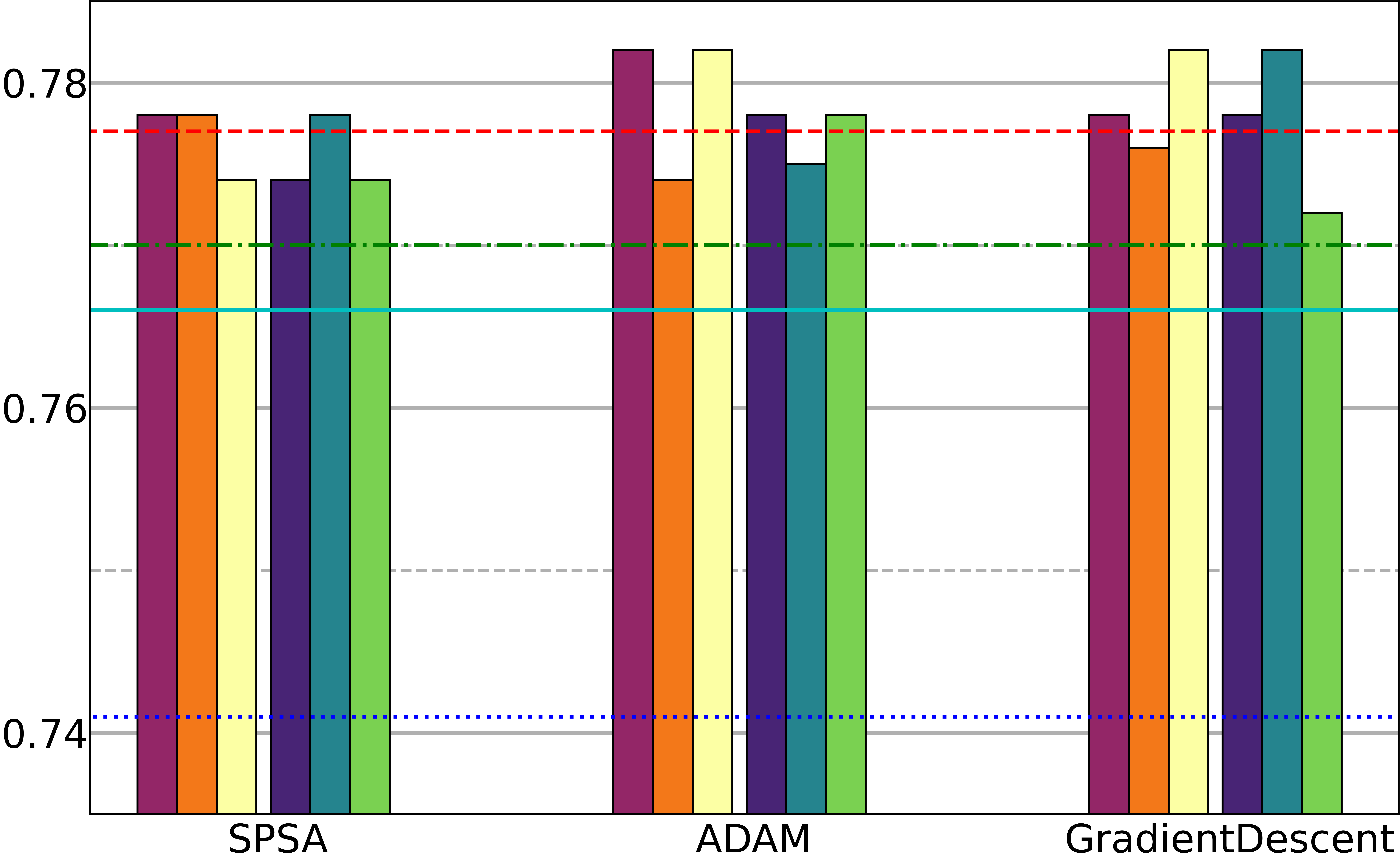}
    \end{minipage}
    \hfill
    \begin{minipage}[t]{0.4\textwidth}
        \centering
        \includegraphics[width=\textwidth]{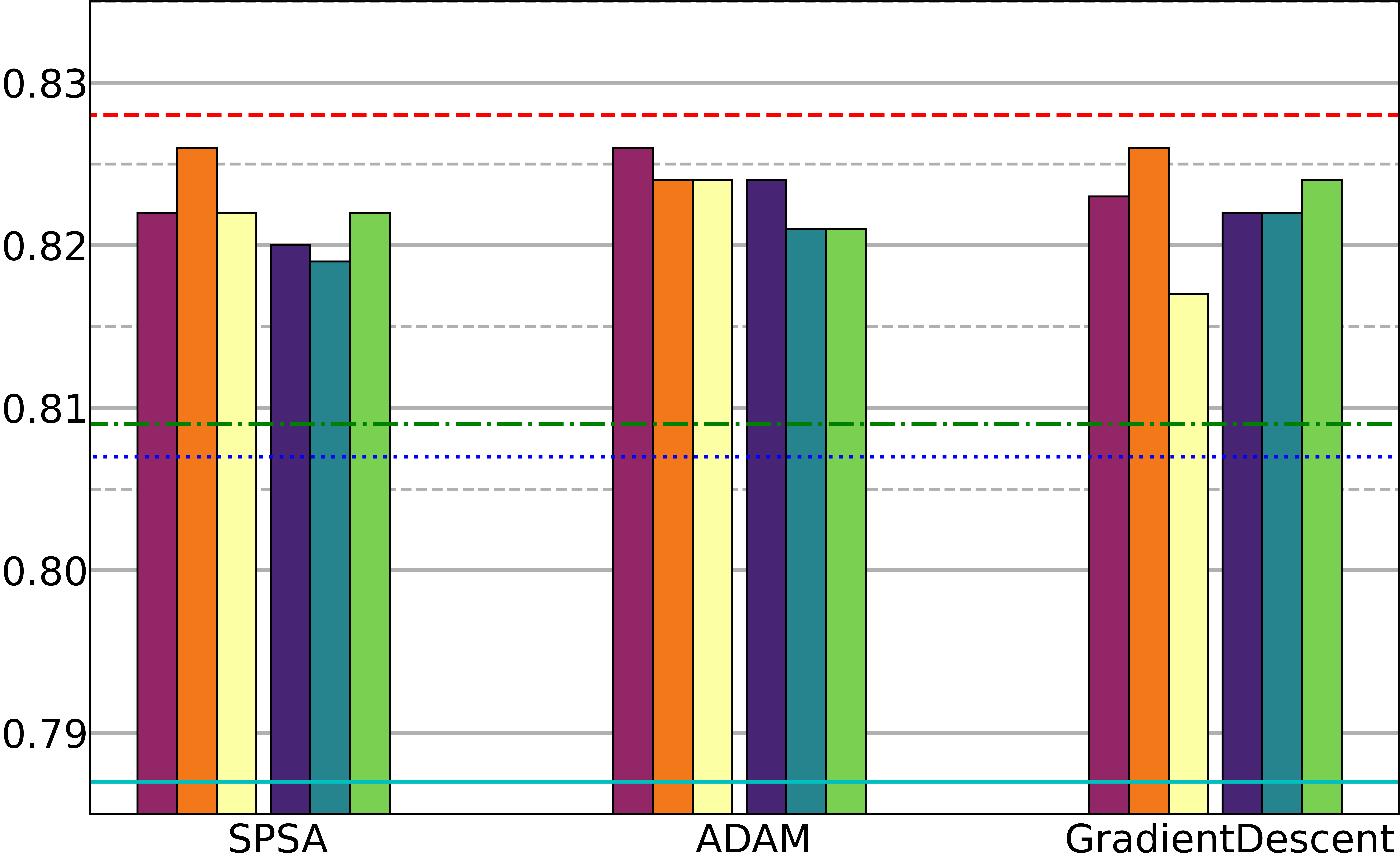}
    \end{minipage}
}    
\resizebox{\textwidth}{!}{       
    \begin{minipage}[t]{0.05\textwidth}
        \centering
        \includegraphics[width=\textwidth]{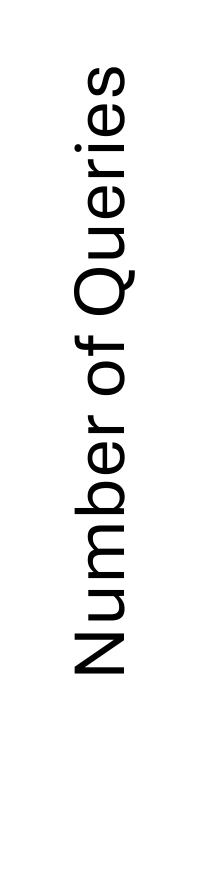}
    \end{minipage}
    \begin{minipage}[t]{0.4\textwidth}
        \centering
        \includegraphics[width=\textwidth]{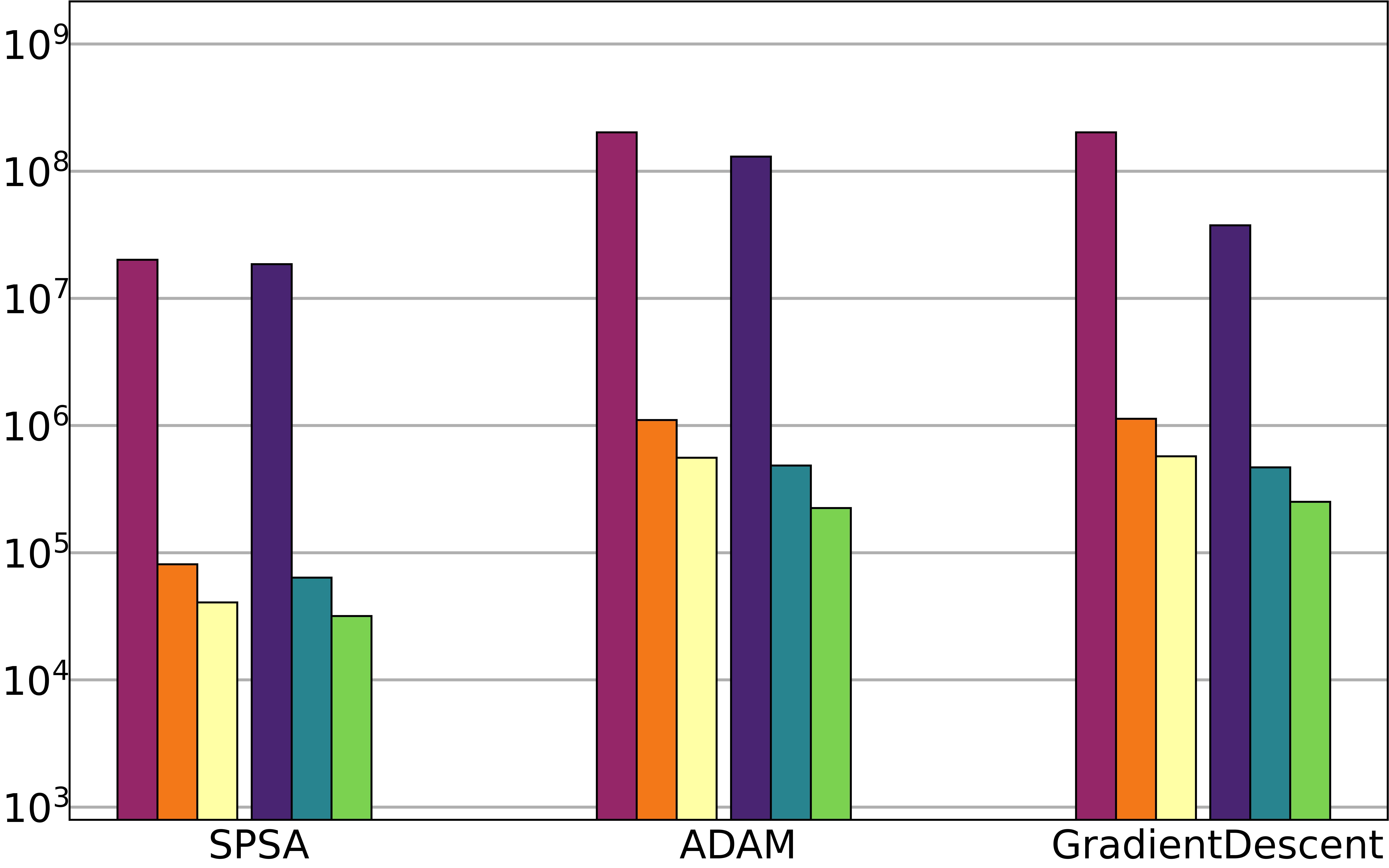}
        
        \caption[]{AF vs D}
    \end{minipage}
    \hfill
    \begin{minipage}[t]{0.4\textwidth}
        \centering
        \includegraphics[width=\textwidth]{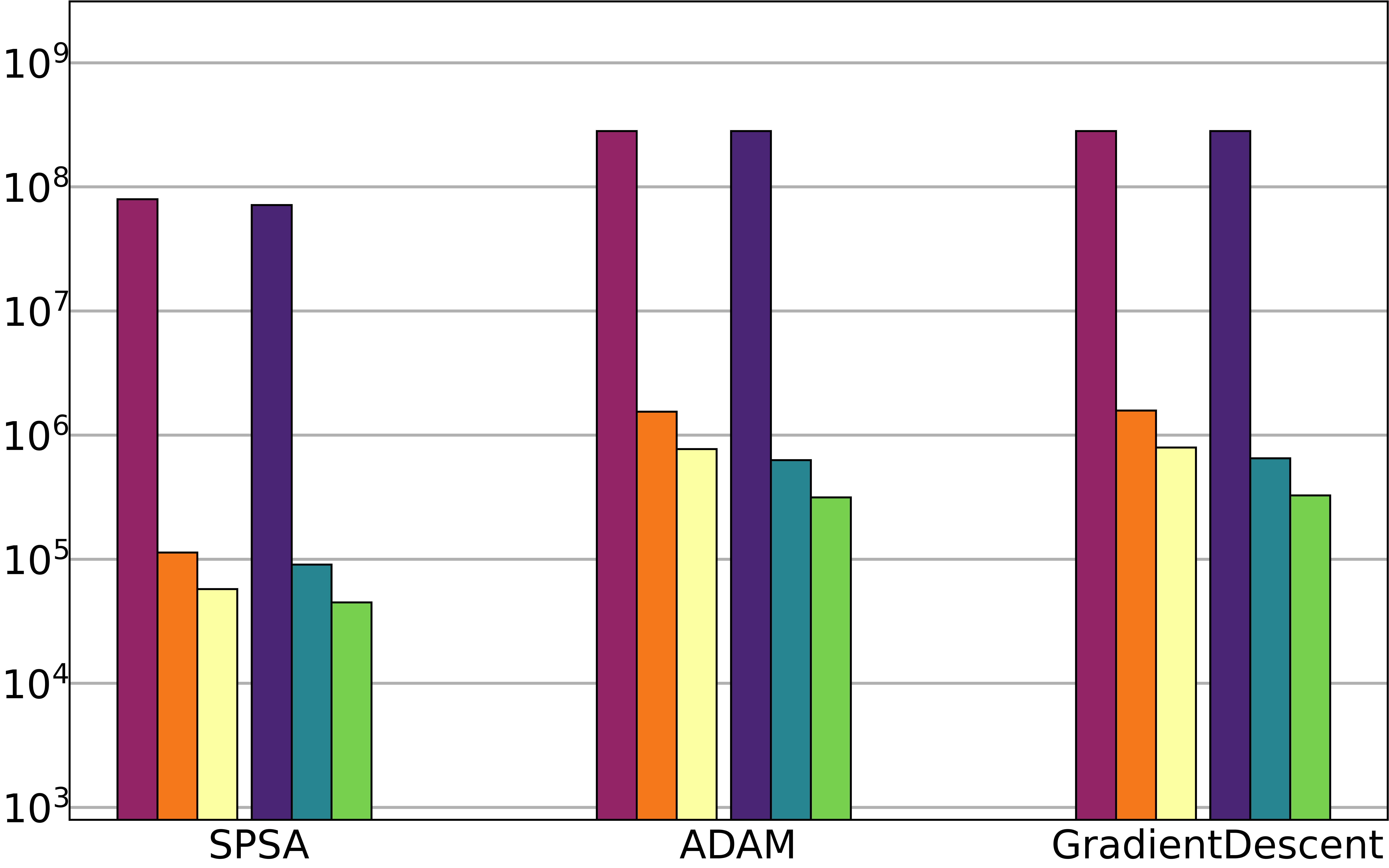}
        \caption[]{NBU vs AFD}
        \vspace{1ex}
    \end{minipage}
}
\endgroup

\begin{minipage}[t]{\textwidth} 
    \caption{Top left and right panels, F1 scores obtained with our method, grouped by optimiser (ADAM, GD and SPSA), using the \textit{full dataset} and for the classification of AF vs D and NBU vs AFD cancer subtypes. Quantum fidelity kernel, best classical kernel and GNN reference values are represented as horizontal lines. Bottom left and right panels: total number of queries for each method, grouped by optimiser. The best results are obtained after a parameter grid search as described in \ref{Apx-Exp}, and full results for PEGASOS QKA is given in \ref{Apx-PGS}}
    \label{BRACS-results-full}
\end{minipage}

\end{figure}
\end{widetext}

It's important to note that the results (shown in Fig. \ref{BRACS-results-full}) are influenced by the inherent limitations of this feature map. Given that the goal of this work is to demonstrate the performance of the sub-sampling method, we chose this feature map not because it is optimal but simply because it is commonly used and readily available. We believe that it may be possible to improve classification accuracy beyond the results presented here by, for example, exploring different feature maps.

The results highlighted in Fig. \ref{BRACS-results-full} are collected to show the performance of the sub-sampling method comparing all the optimisers at once. In the same Figure, we also make the comparison between the classification accuracy of a GNN, classical kernels, PEGASOS kernel alignment \cite{gentinetta2023quantum} and a quantum fidelity kernel without kernel alignment. In the case of the sub-kernel method applied to the real-world problem of cancer subtyping, we observe similar trends to those observed for the synthetic datasets discussed above. The sub-sampling method does not degrade the classification performance,(top panels in Fig. \ref{BRACS-results-full}). It is also evident that the quantum kernel with QKA outperforms both the quantum fidelity kernel without kernel alignment and PEGASOS kernel alignment in terms of classification accuracy (cyan line and blue dotted line in Fig. \ref{BRACS-results-full}). The comparison of the results for the breast cancer data set between the sub-kernel method and the classical GNN approach indicates that the sub-kernel method can achieve comparable results to the classical method.  While the initial findings are promising, they should be interpreted as indicative rather than conclusive.

The advantages of employing the sub-sampling method to accelerate kernel training are also clear (bottom panels in Fig. \ref{BRACS-results-full}). Given the substantial size of the full kernels in these cases (ranging from $\sim$1600 to $\sim$2800 data points), the reduction in queries is even more significant than previously observed, with some instances requiring three orders of magnitude fewer queries. When a kernel of size $m=2796$ is reduced to a sub-kernel of size $k=8$ and $s=1$, the number of queries is reduced by a factor of $1.6\times10^3$. These results provide some evidence that it is possible to choose the parameters $s$ and $k$ as constants independent of $m$. In the case that $m=1566$, we find that choosing $k=16$ and $s=1$ gives a worst case reduction in ROC AUC and F1 score of 1.77\% and 1.02\% respectively. On average ROC AUC increases by 1.16\% and the F1 score decreases by 0.1\% (see Appendix E Table X). Increasing the full kernel size to $m=2796$ gives similarly good results for  the same parameters, $k=16$ and $s=1$. The worst case reduction in ROC and F1 score is 1.18\% and 0.36\% respectively. On average the ROC AUC increases by 0.61\% the F1 score increases by 0.02\% (see Appendix E Table XII). While we can't generalise from these two data points, the fact that we observe that $k$ and $s$ do not need to grow with $m$ in this case is an encouraging sign that the sub-kernel method can provide highly efficient scaling, potentially independent of $m$.

\section{Conclusion}\label{sec12}

This study proposes a novel approach for implementing quantum kernel alignment for Support Vector Classification (SVC), introducing the use of sub-sampling during the quantum training phase. QKA is a method that is capable of improving the performance of quantum kernels. However, QKA comes at a significant cost due to the need to train the kernel. This cost has, until now, undermined the practical usefulness of quantum kernel alignment. However, our results have demonstrated that the sub-sampling method is a promising approach that makes QKA more viable in the near-term by considerably reducing number of circuits required to train the kernel. Importantly we have observed that this reduction in training cost does not necessarily come at the cost of classification accuracy.

While the sub-sampling method has the potential to speed-up the training process, there are no \textit{a priori} guarantees that this will be the case. The actual reduction (or increase) in queries will depend on the choice of sub-kernel size, number of samples per training iteration and the specifics of the dataset. Nonetheless, in practice we have found that, in the cases tested, the speed-up is remarkably robust across different parameters, ansatzes and classical optimisers. The vast majority of instances tested resulted in reductions in number of circuits, even when testing on a real-world dataset. In fact the real-world dataset gave the best results with reductions of up to 3 orders of magnitude. Furthermore, the sub-sampling method introduces a level of stochastic optimisation which could lessen the severity of errors. This may explain why we observe improvements in classification accuracy in several cases when using the sub-sampling method. However, the stochastic nature of the training phase may also cause stability and reproducibility issues.

\begin{acknowledgments}

This work was supported by the Hartree National Centre for Digital Innovation, a collaboration between the Science and Technology Facilities Council and IBM. We thank Stefan Woerner, Francesco Tacchino and Ivano Tavernelli for their valuable inputs. 
The project concept and research proposal was created and developed by Imperial College London and the Royal Brompton and Harefield Hospitals in collaboration with IBM.
We also thank Voica Ana Maria Radescu, Michele Grossi, Chris Burton, Frederik Flöther for their valuable support and input to the initial project development and concept. 

\end{acknowledgments}

\bibliography{main}

\clearpage

\onecolumngrid
\appendix


\section{Details of Experiments}\label{Apx-Exp}

In our numerical experiments, we utilized specific software and dataset settings to ensure reproducibility and precision. The software versions employed were Qiskit version 0.44, Qiskit Machine Learning version 0.6, and scikit-learn Version 1.3.0. For the Synthetic Benchmarking Dataset, the settings were: SPSA with 200 maximum iterations, a learning rate of 0.01, and a perturbation of 0.05; ADAM with 200 maximum iterations, a tolerance of 1e-06, and a learning rate of 0.01; Gradient Descent with 200 maximum iterations, a learning rate of 0.01, a tolerance of 1e-07, and no perturbation. For the COSET Dataset and the Computational Pathology Breast Cancer Dataset, maximum iterations were increased to 400 for all algorithms, keeping other parameters constant.

Using a validation set, which is split from the training set, a grid-search is performed over $C$ and $\gamma$ parameters as well as kernel type (Poly, RFB, Linear), and the best parameters and kernel type are then used to find best Classical Kernel to train the SVC algorithm.
Similarly, a grid-search is performed over different $C$ values for the obtained quantum kernels using the same validation set for both QKA and PEGASOS QKA algorithms.

\section{Results of PEGASOS QKA}\label{Apx-PGS}

\begin{table}[H]
\centering
\caption{Reference Values for NBU vs AFD and AF vs D. Best results of PEGASOS QKA \cite{gentinetta2023quantum} from the grid search as described in Appendix \ref{Apx-Exp}}
\begin{tabular}{|c|c|}
\hline
\multicolumn{2}{|c|}{NBU vs AFD} \\\hline
\textbf{Number of Steps} & \textbf{F1 Score} \\
\hline
10 & 0.602 \\\hline
100 & 0.706 \\\hline
1,000 & 0.765 \\\hline
10,000 & 0.787 \\\hline
100,000 & 0.780 \\\hline
\hline
\multicolumn{2}{|c|}{AF vs D} \\
\hline
\textbf{Number of Steps} & \textbf{F1 Score} \\\hline
10 & 0.737 \\\hline
100 & 0.766 \\\hline
1,000 & 0.763 \\\hline
10,000 & 0.760 \\\hline
100,000 & 0.760 \\\hline
\end{tabular}
\end{table}

\section{Results for Hardware run of Synthetic Dataset with SPSA Optimizer}\label{Apx-hardware}

\begin{figure}[H]
    \centering
    \includegraphics[width=0.2\textwidth]{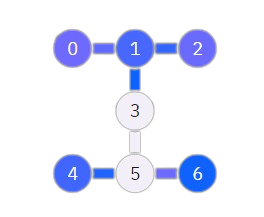}
    \caption{Schematic of the IBM quantum hardware \textit{ibm\_nairobi} device taken at the time of the hardware experiments.}
    \label{fig:hw-schematic}
\end{figure}

The hardware used in this work is IBM's \textit{ibm\_nairobi} with Falcon r5.11H processor and 7 qubits. For qubit connectivity refer to Fig. \ref{fig:hw-schematic}. Qiskit options were set as \textit{initial\_layout=[1,2]}, \textit{transpile=3} and \textit{resillience-level=0} with 100 shots.

\resizebox{\textwidth}{!}{       
    \begin{minipage}[t]{\textwidth}
        \begin{table}[H]
        \centering
        \caption{It has been only run once. Hardware results for SPSA Optimizer on the synthetic benchmarking dataset in section 2.1.1. $k$ is the sub-kernel sample size, and $s$ is the number of samples. The F1 and ROC scores are computed on the full, optimised kernel. Results are ordered by descending $k$ and grouped by ansatze. The runtime is calculated as the total hardware execution time inside an IBM Quantum Session.}
        \begin{tabular}{|c|c|c|c|c|c|c|}
        \hline
        Ansatz & k & s & ROC AUC & F1 & Runtime (s)\\
        \hline
         HE & 96 & 1 & 0.85  & 0.721 & 20155 \\\hline
         HE & 32 & 1 & 0.904 & 0.906 & 2819 \\\hline
         HE & 32 & 8 & 0.94  & 0.889 & 16331 \\\hline
         HE & 16 & 4 & 0.995 & 0.953 & 3590 \\\hline
         HE &  4 & 8 & 0.992 & 0.953 & 2422 \\\hline
         RA & 96 & 1 & 0.86  & 0.8  & 19453 \\\hline
         RA & 32 & 1 & 0.807 & 0.694 & 2864 \\\hline
         RA & 32 & 8 & 0.859 & 0.766 & 15506 \\\hline
         RA & 16 & 4 & 0.891 & 0.812 & 3851 \\\hline
        \end{tabular}
        \label{table:hw}
        \end{table}
    \end{minipage}
}

\section{Loss Graphs}\label{Apx-loss}

\begin{figure}[H]
    \centering
    \centering
    \begin{minipage}[t]{\textwidth}
        \centering
        \includegraphics[width=\textwidth]{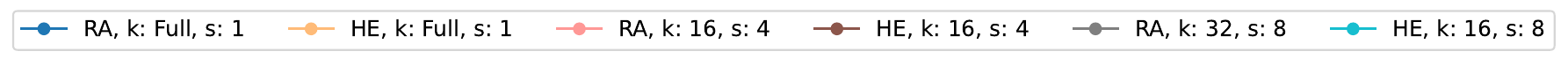}
        \includegraphics[width=\textwidth]{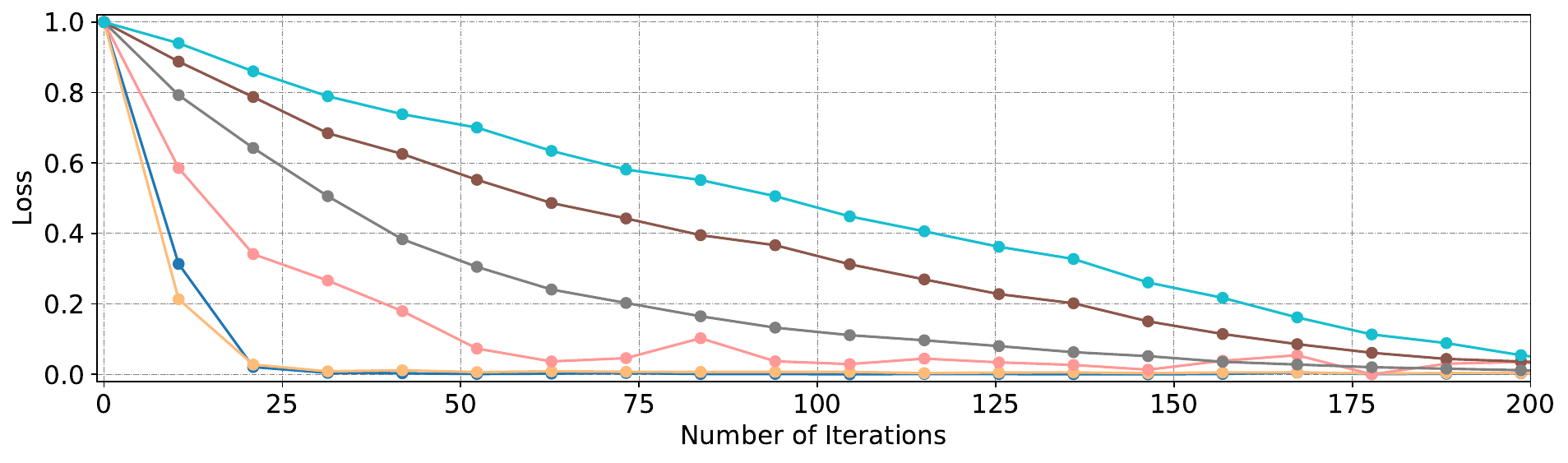}
        \caption{Min-Max normalised loss function for SPSA optimiser on the synthetic benchmarking dataset in section 2.1.1}
        \label{fig:Synthetic-SPSA-LOSS}
    \end{minipage}
\end{figure}

\begin{figure}[H]
    \centering
    \begin{minipage}[t]{\textwidth}
        \centering
        \includegraphics[width=\textwidth]{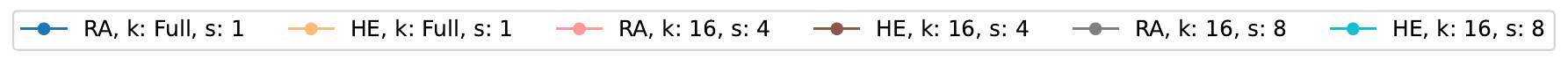}
        \includegraphics[width=\textwidth]{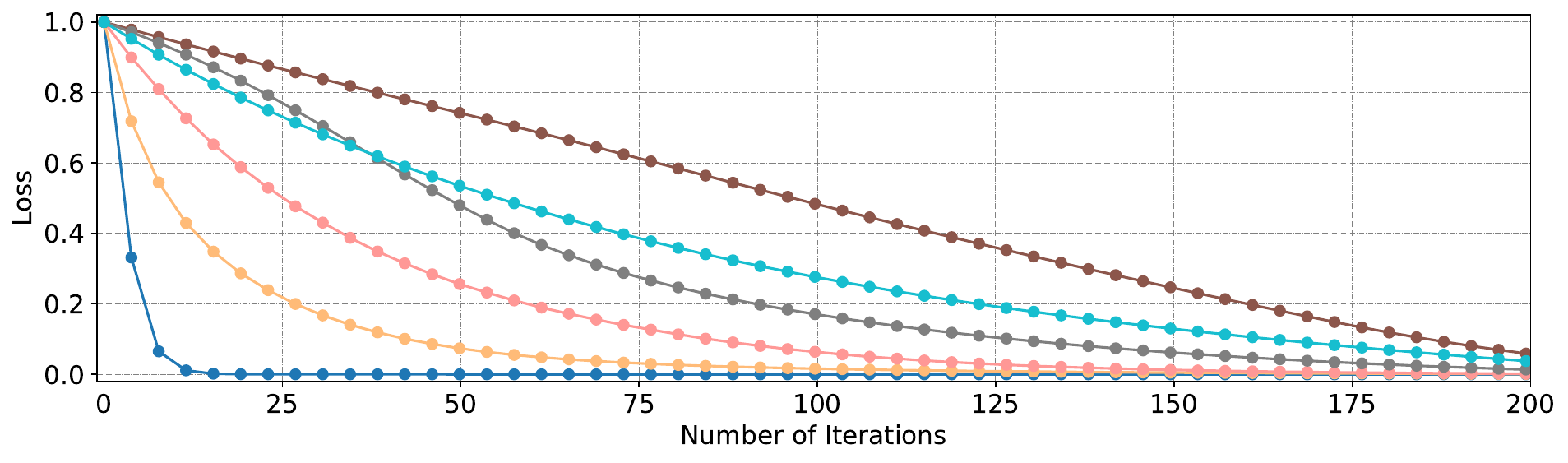}
        \caption{Min-Max normalised loss function for GD optimiser on the synthetic benchmarking dataset in the section 2.1.1}
        \label{fig:Synthetic-GD-LOSS}
    \end{minipage}
\end{figure}

\begin{figure}[H]
    \centering
    \centering
    \begin{minipage}[t]{\textwidth}
        \centering
        \includegraphics[width=\textwidth]{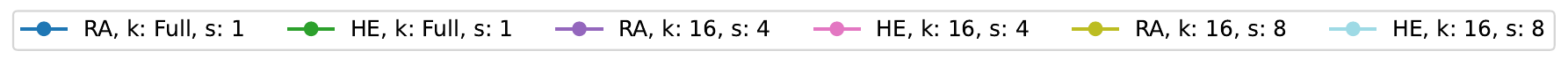}
        \includegraphics[width=\textwidth]{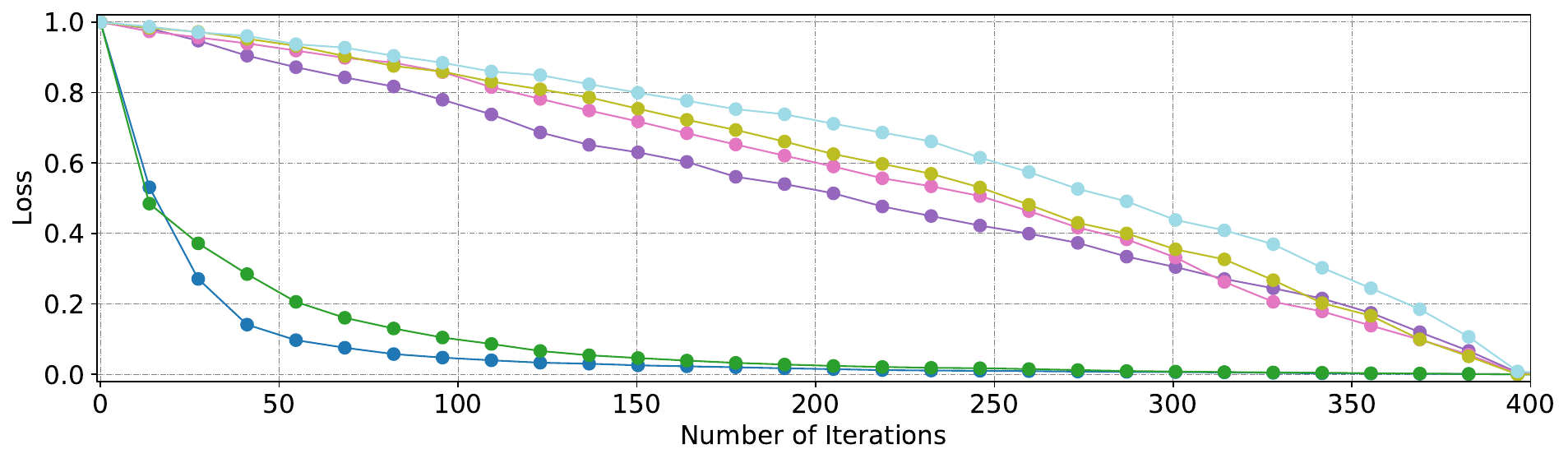}
        \caption{Min-Max normalised loss function for SPSA optimiser on COSETS problem.}
        \label{fig:COSETS-SPSA-LOSS}
    \end{minipage}
\end{figure}

The magnitudes of the loss values computed for different settings can vary considerably due to the use of different kernel sizes, leading to significantly different scales of loss. Direct comparisons of these values are not informative because of this scale variation. To facilitate meaningful comparisons, min-max scaling is applied to each loss graph in Figs. \ref{fig:Synthetic-SPSA-LOSS} - \ref{fig:COSETS-SPSA-LOSS}, standardizing the range of values. The purpose of these plots is to demonstrate that, as expected, the sub-sampling method does result in the loss converging at a slower rate compared to the full kernel method.

\begin{figure}[H]
    \centering
    \begin{minipage}[t]{\textwidth}
        \centering
        \includegraphics[width=\textwidth]{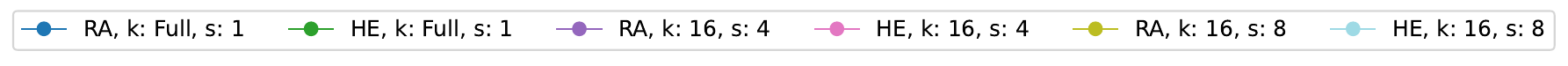}
        \includegraphics[width=\textwidth]{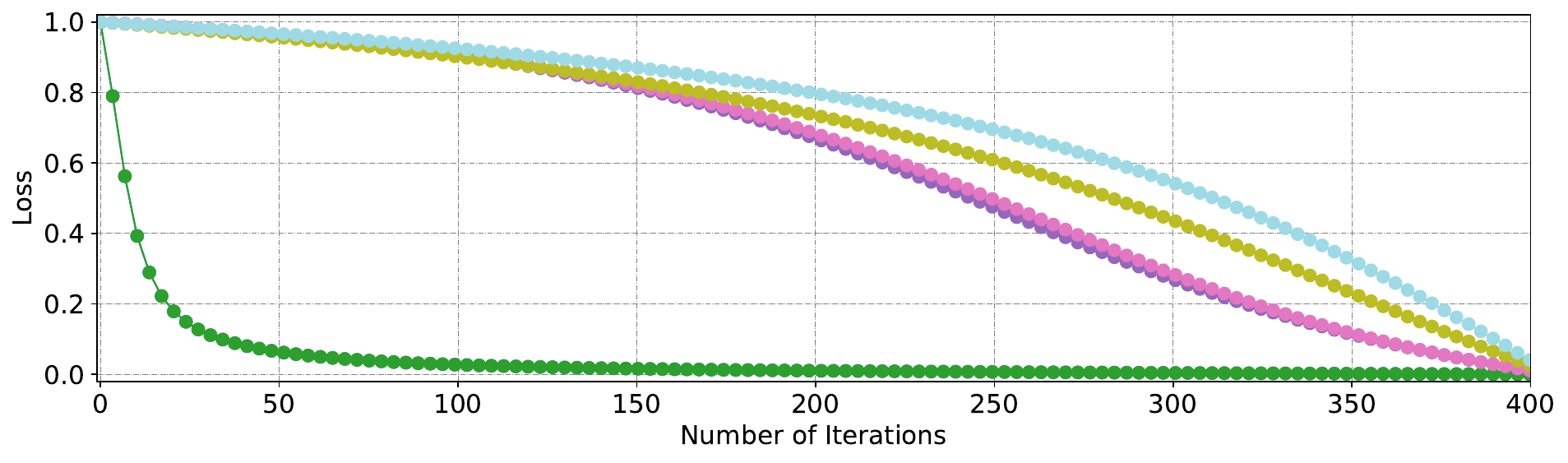}
        \caption{Min-Max normalised loss function for GD optimiser on COSETS problem.}
        \label{fig:COSETS-GD-LOSS}
    \end{minipage}
\end{figure}
\clearpage

\section{Numerical Results}\label{Apx-Num}

\begin{table}[!h]
\centering
\caption{Results for SPSA Optimizer on the synthetic benchmarking dataset in section 2.1.1. $k$ is the sub-kernel sample size, and $s$ is the number of samples. The F1 and ROC scores are computed on the full, optimised kernel. Results are ordered by descending $k$ and grouped by ansatze. Full kernels ($k=m$) are highlighted in bold.}
\resizebox{0.4\textwidth}{!}{
\begin{tabular}{|c|c|c|c|c|c|c|}
\hline
Ansatz & k & s & ROC AUC & F1 & Queries & Speed-up \\
\hline
\textbf{HE} & \textbf{96} & \textbf{1} & \textbf{0.98} & \textbf{0.9312} & \textbf{1377696} & \textbf{1} \\\hline
HE & 32 & 1 & 0.99 & 0.9312 & 378848 & 3.64 \\\hline
HE & 32 & 4 & 0.985 & 0.9312 & 1380320 & 1 \\\hline
HE & 32 & 8 & 0.98 & 0.9312 & 3125216 & 0.44 \\\hline
HE & 16 & 1 & 0.985 & 0.8561 & 105712 & 13.03 \\\hline
HE & 16 & 4 & 0.99 & 0.9312 & 452080 & 3.05 \\\hline
HE & 16 & 8 & 0.985 & 0.9312 & 894960 & 1.54 \\\hline
HE & 8 & 1 & 0.98 & 0.9312 & 14808 & 93.04 \\\hline
HE & 8 & 4 & 0.985 & 0.9312 & 108536 & 12.69 \\\hline
HE & 8 & 8 & 0.985 & 0.9312 & 213752 & 6.45 \\\hline
HE & 4 & 1 & 0.96 & 0.7407 & 3012 & 457.40 \\\hline
HE & 4 & 4 & 0.965 & 0.8667 & 26908 & 51.20 \\\hline
HE & 4 & 8 & 1 & 0.9312 & 54460 & 25.30 \\\hline
\textbf{RA} & \textbf{96} & \textbf{1} & \textbf{0.98} & \textbf{0.9312} & \textbf{852384} & \textbf{1} \\\hline
RA & 32 & 1 & 0.985 & 0.9312 & 33760 & 25.25 \\\hline
RA & 32 & 4 & 0.985 & 0.9312 & 1558496 & 0.55 \\\hline
RA & 32 & 8 & 0.990 & 0.9312 & 3063776 & 0.28 \\\hline
RA & 16 & 1 & 0.985 & 0.9312 & 87024 & 9.79 \\\hline
RA & 16 & 4 & 0.985 & 0.9312 & 260784 & 3.26 \\\hline
RA & 16 & 8 & 0.89 & 0.8244 & 919536 & 0.93 \\\hline
RA & 8 & 1 & 0.98 & 0.8624 & 27960 & 30.49 \\\hline
RA & 8 & 4 & 0.97 & 0.8986 & 101496 & 8.4 \\\hline
RA & 8 & 8 & 0.975 & 0.8986 & 227064 & 3.75 \\\hline
RA & 4 & 1 & 0.905 & 0.8244 & 4972 & 171.44 \\\hline
RA & 4 & 4 & 0.97 & 0.8986 & 28636 & 29.77 \\\hline
RA & 4 & 8 & 0.865 & 0.7634 & 56700 & 15.03 \\\hline

\end{tabular}}
\label{apx-spsa-zz}
\end{table}

\begin{table}[!]
\centering
\caption{Results for ADAM Optimizer on the synthetic benchmarking dataset in section 2.1.1. $k$ is the sub-kernel sample size, and $s$ is the number of samples. The F1 and ROC scores are computed on the full, optimised kernel. Results are ordered by descending $k$ and grouped by ansatze. Full kernels ($k=m$) are highlighted in bold.}
\resizebox{0.4\textwidth}{!}{
\begin{tabular}{|c|c|c|c|c|c|c|}
\hline
Ansatz & k & s & ROC AUC & F1 & Queries & Speed-up \\
\hline
\textbf{HE} & \textbf{96} & \textbf{1} & \textbf{0.985} & \textbf{0.9312} & \textbf{6114720} & \textbf{1} \\\hline
HE & 32 & 1 & 0.97 & 0.7407 & 2157024 & 2.83 \\\hline
HE & 32 & 4 & 0.915 & 0.8244 & 3012576 & 2.03 \\\hline
HE & 32 & 8 & 0.96 & 0.8624 & 491488 & 12.44 \\\hline
HE & 16 & 1 & 0.95 & 0.8561 & 111472 & 54.85 \\\hline
HE & 16 & 4 & 0.955 & 0.8624 & 1893872 & 3.23 \\\hline
HE & 16 & 8 & 0.98 & 0.9312 & 902128 & 6.78 \\\hline
HE & 8 & 1 & 0.98 & 0.9312 & 344 & 17775.35 \\\hline
HE & 8 & 4 & 0.975 & 0.9662 & 689400 & 8.87 \\\hline
HE & 8 & 8 & 0.975 & 0.8624 & 158456 & 38.59 \\\hline
HE & 4 & 1 & 0.915 & 0.7937 & 60 & 101912 \\\hline
HE & 4 & 4 & 0.985 & 0.9312 & 148220 & 41.25 \\\hline
HE & 4 & 8 & 0.955 & 0.8624 & 344956 & 17.73 \\\hline
\textbf{RA} & \textbf{96} & \textbf{1} & \textbf{0.91} & \textbf{0.5333} & \textbf{350112} & \textbf{1} \\\hline
RA & 32 & 1 & 0.94 & 0.8986 & 56800 & 6.16 \\\hline
RA & 32 & 4 & 0.965 & 0.8244 & 88032 & 3.98 \\\hline
RA & 32 & 8 & 0.905 & 0.8 & 9957344 & 0.04 \\\hline
RA & 16 & 1 & 0.87 & 0.6936 & 66416 & 5.27 \\\hline
RA & 16 & 4 & 0.985 & 0.8624 & 68080 & 5.14 \\\hline
RA & 16 & 8 & 0.975 & 0.8624 & 984048 & 0.36 \\\hline
RA & 8 & 1 & 0.85 & 0.8000 & 184 & 1902.78 \\\hline
RA & 8 & 4 & 0.915 & 0.8148 & 383736 & 0.91 \\\hline
RA & 8 & 8 & 0.87 & 0.7937 & 768248 & 0.46 \\\hline
RA & 4 & 1 & 0.895 & 0.8244 & 116 & 3018.21 \\\hline
RA & 4 & 4 & 0.97 & 0.9662 & 32060 & 10.92 \\\hline
RA & 4 & 8 & 0.93 & 0.8624 & 172732 & 2.03 \\\hline

\end{tabular}}
\end{table}

\begin{table}[!]
\centering
\caption{Results for GD Optimizer on the synthetic benchmarking dataset in section 2.1.1. $k$ is the sub-kernel sample size, and $s$ is the number of samples. The F1 and ROC scores are computed on the full, optimised kernel. Results are ordered by descending $k$ and grouped by ansatze. Full $k$ are highlighted in bold.}
\resizebox{0.4\textwidth}{!}{
\begin{tabular}{|c|c|c|c|c|c|c|}
\hline
Ansatz & k & s & ROC AUC & F1 & Queries & Speed-up \\
\hline
\textbf{HE} & \textbf{96} & \textbf{1} & \textbf{0.98} & \textbf{0.9312} & \textbf{6437280} & \textbf{1} \\\hline
HE & 32 & 1 & 0.98 & 0.9312 & 1163232 & 5.53 \\\hline
HE & 32 & 4 & 0.985 & 0.9312 & 6948832 & 0.93 \\\hline
HE & 32 & 8 & 0.98 & 0.9312 & 13897696 & 0.46 \\\hline
HE & 16 & 1 & 0.98 & 0.9312 & 323952 & 19.87 \\\hline
HE & 16 & 4 & 0.99 & 0.8947 & 2047984 & 3.14 \\\hline
HE & 16 & 8 & 0.98 & 0.9312 & 4088816 & 1.57 \\\hline
HE & 8 & 1 & 0.97 & 0.8624 & 2520 & 2554.48 \\\hline
HE & 8 & 4 & 0.965 & 0.8947 & 511992 & 12.57 \\\hline
HE & 8 & 8 & 0.965 & 0.8986 & 1023224 & 6.29 \\\hline
HE & 4 & 1 & 0.975 & 0.8624 & 31876 & 201.95 \\\hline
HE & 4 & 4 & 0.975 & 0.8561 & 68668 & 93.74 \\\hline
HE & 4 & 8 & 0.85 & 0.7542 & 189180 & 34.03 \\\hline
\textbf{RA} & \textbf{96} & \textbf{1} & \textbf{0.98} & \textbf{0.9312} & \textbf{2386848} & \textbf{1} \\\hline
RA & 32 & 1 & 0.915 & 0.831 & 1036768 & 2.3 \\\hline
RA & 32 & 4 & 0.985 & 0.9312 & 4171744 & 0.57 \\\hline
RA & 32 & 8 & 0.85 & 0.7333 & 8347616 & 0.29 \\\hline
RA & 16 & 1 & 0.865 & 0.5333 & 2416 & 987.93 \\\hline
RA & 16 & 4 & 0.985 & 0.9312 & 1226736 & 1.95 \\\hline
RA & 16 & 8 & 0.99 & 0.9312 & 2455536 & 0.97 \\\hline
RA & 8 & 1 & 0.945 & 0.8667 & 2520 & 947.16 \\\hline
RA & 8 & 4 & 0.93 & 0.831 & 192376 & 12.41 \\\hline
RA & 8 & 8 & 0.97 & 0.8624 & 614392 & 3.88 \\\hline
RA & 4 & 1 & 0.81 & 0.7249 & 6244 & 382.26 \\\hline
RA & 4 & 4 & 0.985 & 0.8624 & 76604 & 31.16 \\\hline
RA & 4 & 8 & 0.915 & 0.7634 & 13564 & 175.97 \\\hline

\end{tabular}}
\end{table}

\begin{table}[!]
\centering
\caption{Results for SPSA Optimizer on the COSET dataset in section 2.1.2. $k$ is the sub-kernel sample size, and $s$ is the number of samples. The F1 and ROC scores are computed on the full, optimised kernel. Results are ordered by descending $k$ and grouped by ansatze. Full kernels ($k=m$) are highlighted in bold.}
\resizebox{0.4\textwidth}{!}{
\begin{tabular}{|c|c|c|c|c|c|c|}
\hline
Ansatz & k & s & ROC AUC & F1 & Queries & Speed-up \\
\hline
\textbf{HE} & \textbf{96} & \textbf{1} & \textbf{1} & \textbf{1} & \textbf{1396128} & \textbf{1} \\\hline
HE & 32 & 1 & 1 & 1 & 382944 & 3.65 \\\hline
HE & 32 & 4 & 1 & 1 & 1542112 & 0.91 \\\hline
HE & 32 & 8 & 1 & 1 & 3125216 & 0.45 \\\hline
HE & 16 & 1 & 1 & 1 & 113136 & 12.34 \\\hline
HE & 16 & 4 & 1 & 1 & 455152 & 3.07 \\\hline
HE & 16 & 8 & 1 & 0.9687 & 901104 & 1.55 \\\hline
HE & 8 & 1 & 0.9569 & 0.811 & 11736 & 118.96 \\\hline
HE & 8 & 4 & 0.8510 & 0.75 & 96504 & 14.47 \\\hline
HE & 8 & 8 & 0.7765 & 0.75 & 228600 & 6.11 \\\hline
HE & 4 & 1 & 0.7412 & 0.75 & 1668 & 837.01 \\\hline
HE & 4 & 4 & 0.7137 & 0.75 & 18940 & 73.71 \\\hline
HE & 4 & 8 & 0.6941 & 0.7179 & 35452 & 39.38 \\\hline
\textbf{RA} & \textbf{96} & \textbf{1} & \textbf{1} & \textbf{1} & \textbf{1428384} & \textbf{1} \\\hline
RA & 32 & 1 & 1 & 1 & 385504 & 3.71 \\\hline
RA & 32 & 4 & 1 & 1 & 1507296 & 0.95 \\\hline
RA & 32 & 8 & 1 & 1 & 3084256 & 0.46 \\\hline
RA & 16 & 1 & 1 & 1 & 106224 & 13.45 \\\hline
RA & 16 & 4 & 1 & 1 & 446448 & 3.20 \\\hline
RA & 16 & 8 & 1 & 0.9687 & 917488 & 1.56 \\\hline
RA & 8 & 1 & 0.6157 & 0.6161 & 22040 & 64.81 \\\hline
RA & 8 & 4 & 0.9255 & 0.811 & 115064 & 12.41 \\\hline
RA & 8 & 8 & 0.9137 & 0.811 & 229880 & 6.21 \\\hline
RA & 4 & 1 & 0.7882 & 0.75 & 2972 & 480.61 \\\hline
RA & 4 & 4 & 0.6980 & 0.7179 & 27516 & 51.91 \\\hline
RA & 4 & 8 & 0.7686 & 0.75 & 33660 & 42.44 \\\hline

\end{tabular}}
\end{table}

\begin{table}[!]
\centering
\caption{Results for ADAM Optimizer on the COSET dataset in section 2.1.2. $k$ is the sub-kernel sample size, and $s$ is the number of samples. The F1 and ROC scores are computed on the full, optimised kernel. Results are ordered by descending $k$ and grouped by ansatze. Full kernels ($k=m$) are highlighted in bold.}
\label{apx-spsa-coset}
\resizebox{0.4\textwidth}{!}{
\begin{tabular}{|c|c|c|c|c|c|c|}
\hline
Ansatz & k & s & ROC AUC & F1 & Queries & Speed-up \\
\hline
\textbf{HE} & \textbf{96} & \textbf{1} & \textbf{1} & \textbf{1} & \textbf{7741344} & \textbf{1} \\\hline
HE & 32 & 1 & 0.9137 & 0.811 & 2088928 & 3.71 \\\hline
HE & 32 & 4 & 0.9765 & 0.811 & 8173536 & 0.95 \\\hline
HE & 32 & 8 & 0.9804 & 0.9375 & 5533664 & 1.40 \\\hline
HE & 16 & 1 & 0.9176 & 0.8125 & 614512 & 12.60 \\\hline
HE & 16 & 4 & 0.7176 & 0.7179 & 1090544 & 7.10 \\\hline
HE & 16 & 8 & 0.8118 & 0.75 & 4916208 & 1.57 \\\hline
HE & 8 & 1 & 1 & 1 & 153272 & 50.51 \\\hline
HE & 8 & 4 & 0.6941 & 0.7179 & 18936 & 408.82 \\\hline
HE & 8 & 8 & 0.7255 & 0.75 & 146424 & 52.87 \\\hline
HE & 4 & 1 & 0.9843 & 0.906 & 38268 & 202.29 \\\hline
HE & 4 & 4 & 0.9725 & 0.874 & 153500 & 50.43 \\\hline
HE & 4 & 8 & 1 & 1 & 306812 & 25.23 \\\hline
\textbf{RA} & \textbf{96} & \textbf{1} & \textbf{1} & \textbf{1} & \textbf{7741344} & \textbf{1} \\\hline
RA & 32 & 1 & 0.7647 & 0.75 & 1883616 & 4.11 \\\hline
RA & 32 & 4 & 0.8980 & 0.811 & 8355808 & 0.93 \\\hline
RA & 32 & 8 & 0.8627 & 0.7806 & 16711648 & 0.46 \\\hline
RA & 16 & 1 & 0.9961 & 0.8414 & 613872 & 12.61 \\\hline
RA & 16 & 4 & 0.8118 & 0.75 & 2391024 & 3.24 \\\hline
RA & 16 & 8 & 0.7216 & 0.75 & 3724272 & 2.08 \\\hline
RA & 8 & 1 & 1 & 1 & 153528 & 50.42 \\\hline
RA & 8 & 4 & 0.7373 & 0.75 & 613496 & 12.62 \\\hline
RA & 8 & 8 & 0.8275 & 0.75 & 1229048 & 6.3 \\\hline
RA & 4 & 1 & 1 & 1 & 38388 & 201.66 \\\hline
RA & 4 & 4 & 1 & 1 & 153468 & 50.44 \\\hline
RA & 4 & 8 & 0.7294 & 0.75 & 137980 & 56.1 \\\hline

\end{tabular}}
\end{table}

\begin{table}[!]
\centering
\caption{Results for GD Optimizer on the COSET dataset in section 2.1.2. $k$ is the sub-kernel sample size, and $s$ is the number of samples. The F1 and ROC scores are computed on the full, optimised kernel. Results are ordered by descending $k$ and grouped by ansatze. Full kernels ($k=m$) are highlighted in bold.}
\resizebox{0.4\textwidth}{!}{
\begin{tabular}{|c|c|c|c|c|c|c|}
\hline
Ansatz & k & s & ROC AUC & F1 & Queries & Speed-up \\
\hline
\textbf{HE} & \textbf{96} & \textbf{1} & \textbf{1} & \textbf{1} & \textbf{5796768} & \textbf{1} \\\hline
HE & 32 & 1 & 1 & 1 & 1564640 & 3.7 \\\hline
HE & 32 & 4 & 1 & 1 & 6262752 & 0.93 \\\hline
HE & 32 & 8 & 1 & 1 & 12525536 & 0.46 \\\hline
HE & 16 & 1 & 1 & 1 & 403184 & 14.38 \\\hline
HE & 16 & 4 & 1 & 1 & 1842160 & 3.15 \\\hline
HE & 16 & 8 & 1 & 1 & 3684336 & 1.57 \\\hline
HE & 8 & 1 & 0.4902 & 0.5795 & 4408 & 1315.06 \\\hline
HE & 8 & 4 & 1 & 1 & 425976 & 13.61 \\\hline
HE & 8 & 8 & 0.7451 & 0.685 & 921080 & 6.29 \\\hline
HE & 4 & 1 & 0.5804 & 0.5795 & 6036 & 960.37 \\\hline
HE & 4 & 4 & 1 & 0.9688 & 75004 & 77.29 \\\hline
HE & 4 & 8 & 1 & 1 & 104700 & 55.37 \\\hline
\textbf{RA} & \textbf{96} & \textbf{1} & \textbf{1} & \textbf{1} & \textbf{5796768} & \textbf{1} \\\hline
RA & 32 & 1 & 1 & 1 & 774624 & 7.48 \\\hline
RA & 32 & 4 & 1 & 1 & 6262752 & 0.93 \\\hline
RA & 32 & 8 & 1 & 1 & 12525536 & 0.46 \\\hline
RA & 16 & 1 & 0.9529 & 0.875 & 311664 & 18.6 \\\hline
RA & 16 & 4 & 1 & 1 & 1842160 & 3.15 \\\hline
RA & 16 & 8 & 1 & 1 & 3684336 & 1.57 \\\hline
RA & 8 & 1 & 0.7843 & 0.6161 & 28792 & 201.33 \\\hline
RA & 8 & 4 & 0.9647 & 0.906 & 118136 & 49.07 \\\hline
RA & 8 & 8 & 0.7569 & 0.75 & 921336 & 6.29 \\\hline
RA & 4 & 1 & 1 & 0.874 & 22220 & 260.88 \\\hline
RA & 4 & 4 & 1 & 0.906 & 107164 & 54.09 \\\hline
RA & 4 & 8 & 0.702 & 0.6161 & 198268 & 29.24 \\\hline

\end{tabular}}
\end{table}
\clearpage

\begin{table}[!]
\centering
\caption{Reference Values for AF vs D. The best Classical Kernel Represents the result of grid search as described in Appendix \ref{Apx-Exp}. Classical NN Reference is the result from \cite{pati2022hierarchical}}
\resizebox{0.35\textwidth}{!}{
\begin{tabular}{|c|c|c|}
\hline
 & ROC & F1 \\
\hline
Classical NN Reference & 0.818 & 0.777 \\\hline
Best Classical Kernel & 0.802 & 0.77 \\\hline
Quantum Fidelity Kernel & 0.764 & 0.741 \\\hline
\end{tabular}}
\label{Apx-ref-AFvD}
\end{table}

\begin{table}[!]
\centering
\caption{Results for AF vs D dataset in section 2.1.3. Grouped by optimiser (ADAM, GD and SPSA) $k$ is the sub-kernel sample size, and $s$ is the number of samples. The F1 and ROC scores are computed on the full, optimised kernel. Results are ordered by descending $k$ and grouped by ansatze. Full kernels ($k=m$) are highlighted in bold.}
\resizebox{0.4\textwidth}{!}{
\begin{tabular}{|c|c|c|c|c|c|c|c|}
\hline
Optimizer & Ansatz & k & s & ROC AUC & F1 & Queries & Speed-up \\
\hline
\textbf{SPSA}  &  \textbf{HE}  &  \textbf{1566}  &  \textbf{1}  &  \textbf{0.793}  &  \textbf{0.778}  & \textbf{2.01 x} $\mathbf{10^8}$   &   \textbf{1} \\\hline
SPSA  &  HE  &  16  &  1  &  0.786  &  0.778  & 8.10 x $10^5$   &   248.02 \\\hline
SPSA  &  HE  &  16  &  4  &  0.793  &  0.778  & 0.81 x $10^6$   &   61.61 \\\hline
SPSA  &  HE  &  8  &  1  &  0.774  &  0.774  & 4.06 x $10^5$   &   494.25 \\\hline
SPSA  &  HE  &  8  &  4  &  0.788  &  0.781  & 0.41 x $10^6$   &   127.57 \\\hline
\textbf{SPSA}  &  \textbf{RA}  &  \textbf{1566}  &  \textbf{1}  &  \textbf{0.745}  &  \textbf{0.774}  & \textbf{1.85 x} $\mathbf{10^8}$   &   \textbf{1} \\\hline
SPSA  &  RA  &  16  &  1  &  0.779  &  0.778  & 6.37 x $10^5$   &   291.11 \\\hline
SPSA  &  RA  &  16  &  4  &  0.789  &  0.763  & 0.64 x $10^6$   &   67.45 \\\hline
SPSA  &  RA  &  8  &  1  &  0.775  &  0.774  & 3.18 x $10^5$   &   583.09 \\\hline
SPSA  &  RA  &  8  &  4  &  0.773  &  0.763  & 0.32 x $10^6$   &   138 \\\hline
\textbf{ADAM}  &  \textbf{HE}  &  \textbf{1566}  &  \textbf{1}  &  \textbf{0.791}  &  \textbf{0.782}  & \textbf{2.02 x} $\mathbf{10^9}$   &   \textbf{1} \\\hline
ADAM  &  HE  &  16  &  1  &  0.777  &  0.774  & 1.1 x $10^7$   &   182.47 \\\hline
ADAM  &  HE  &  16  &  4  &  0.787  &  0.774  & 1.1 x $10^7$   &   45.65 \\\hline
ADAM  &  HE  &  8  &  1  &  0.789  &  0.782  & 5.58 x $10^6$   &   361.05 \\\hline
ADAM  &  HE  &  8  &  4  &  0.784  &  0.767  & 0.56 x $10^7$   &   89.30 \\\hline
\textbf{ADAM}  &  \textbf{RA}  &  \textbf{1566}  &  \textbf{1}  &  \textbf{0.774}  &  \textbf{0.778}  & \textbf{1.30 x} $\mathbf{10^9}$   &   \textbf{1} \\\hline
ADAM  &  RA  &  16  &  1  &  0.791  &  0.775  & 4.85 x $10^6$   &   268.40 \\\hline
ADAM  &  RA  &  16  &  4  &  0.775  &  0.774  & 0.48 x $10^7$   &   62.98 \\\hline
ADAM  &  RA  &  8  &  1  &  0.774  &  0.778  & 2.24 x $10^6$   &   580.27 \\\hline
ADAM  &  RA  &  8  &  4  &  0.770  &  0.762  & 2.24 x $10^6$   &   146.58 \\\hline
\textbf{GD}  &  \textbf{HE}  &  \textbf{1566}  &  \textbf{1}  &  \textbf{0.775}  &  \textbf{0.778}  & \textbf{2.02 x} $\mathbf{10^9}$   &   \textbf{1} \\\hline
GD  &  HE  &  16  &  1  &  0.789  &  0.776  & 1.13 x $10^7$   &   178.41 \\\hline
GD  &  HE  &  16  &  4  &  0.782  &  0.782  & 1.13 x $10^7$   &   42.28 \\\hline
GD  &  HE  &  8  &  1  &  0.751  &  0.782  & 5.73 x $10^6$   &   351.83 \\\hline
GD  &  HE  &  8  &  4  &  0.775  &  0.782  & 0.57 x $10^7$   &   88.20 \\\hline
\textbf{GD}  &  \textbf{RA}  &  \textbf{1566}  &  \textbf{1}  &  \textbf{0.783}  &  \textbf{0.778}  & \textbf{3.75 x} $\mathbf{10^8}$   &   \textbf{1} \\\hline
GD  &  RA  &  16  &  1  &  0.791  &  0.782  & 4.68 x $10^6$   &   80.06 \\\hline
GD  &  RA  &  16  &  4  &  0.783  &  0.77  & 0.47 x $10^7$   &   19.76 \\\hline
GD  &  RA  &  8  &  1  &  0.767  &  0.772  & 2.51 x $10^6$   &   149.21 \\\hline
GD  &  RA  &  8  &  4  &  0.795  &  0.778  & 2.51 x $10^6$   &   37.56 \\\hline
\end{tabular}}
\end{table}
\clearpage

\begin{table}[!]
\centering
\caption{Reference Values for NBU vs AFD. The best Classical Kernel Represents result of grid search as described in Appendix \ref{Apx-Exp}. Classical NN Reference is the result from \cite{pati2022hierarchical}}
\resizebox{0.35\textwidth}{!}{
\begin{tabular}{|c|c|c|}
\hline
 & ROC & F1 \\
\hline
Classical NN Reference & 0.868 & 0.828 \\\hline
Best Classical Kernel & 0.859 & 0.809 \\\hline
Quantum Fidelity Kernel & 0.850 & 0.807 \\\hline
\end{tabular}}
\end{table}

\begin{table}[!]
\centering
\caption{Results for NBU vs AFD dataset in section 2.1.3. Grouped by optimiser (ADAM, GD and SPSA) $k$ is the sub-kernel sample size, and $s$ is the number of samples. The F1 and ROC scores are computed on the full, optimised kernel. Results are ordered by descending $k$ and grouped by ansatze. Full kernels ($k=m$) are highlighted in bold.}
\label{Apx-NBUvAFD-full}
\resizebox{0.5\textwidth}{!}{
\begin{tabular}{|c|c|c|c|c|c|c|c|}
\hline
Optimizer & Ansatz & k & s & ROC AUC & F1 & Queries & Speed-up \\
\hline
\textbf{SPSA}  &  \textbf{HE}  &  \textbf{2796}  &  \textbf{1}  &  \textbf{0.844}  &  \textbf{0.822}  & \textbf{7.93 x} $\mathbf{10^8}$   &   \textbf{1} \\\hline
SPSA  &  HE  &  16  &  1  &  0.834  &  0.826  & 1.13 x $10^6$   &   702.02 \\\hline
SPSA  &  HE  &  16  &  4  &  0.843  &  0.828  & 4.03 x $10^6$   &   174.29 \\\hline
SPSA  &  HE  &  8  &  1  &  0.833  &  0.822  & 5.74 x $10^5$   &   1382.46 \\\hline
SPSA  &  HE  &  8  &  4  &  0.836  &  0.828  & 0.57 x $10^6$   &   350.59 \\\hline
\textbf{SPSA}  &  \textbf{RA}  &  \textbf{2796}  &  \textbf{1}  &  \textbf{0.83}  &  \textbf{0.820}  & \textbf{7.14 x} $\mathbf{10^8}$   &   \textbf{1} \\\hline
SPSA  &  RA  &  16  &  1  &  0.837  &  0.819  & 9.03 x $10^5$   &   790.26 \\\hline
SPSA  &  RA  &  16  &  4  &  0.833  &  0.82  & 0.90 x $10^6$   &   194.75 \\\hline
SPSA  &  RA  &  8  &  1  &  0.841  &  0.822  & 4.48 x $10^5$   &   1593.03 \\\hline
SPSA  &  RA  &  8  &  4  &  0.831  &  0.821  & 0.45 x $10^6$   &   370.98 \\\hline
\textbf{ADAM}  &  \textbf{HE}  &  \textbf{2796}  &  \textbf{1}  &  \textbf{0.834}  &  \textbf{0.826}  & \textbf{2.82 x} $\mathbf{10^9}$   &   \textbf{1} \\\hline
ADAM  &  HE  &  16  &  1  &  0.833  &  0.824  & 1.54 x $10^7$   &   182.64 \\\hline
ADAM  &  HE  &  16  &  4  &  0.85  &  0.823  & 1.54 x $10^7$   &   43.07 \\\hline
ADAM  &  HE  &  8  &  1  &  0.826  &  0.824  & 7.71 x $10^6$   &   365.04 \\\hline
ADAM  &  HE  &  8  &  4  &  0.840  &  0.824  & 0.77 x $10^7$   &   93.90 \\\hline
\textbf{ADAM}  &  \textbf{RA}  &  \textbf{2796}  &  \textbf{1}  &  \textbf{0.83}  &  \textbf{0.824}  & \textbf{2.82 x} $\mathbf{10^9}$   &   \textbf{1} \\\hline
ADAM  &  RA  &  16  &  1  &  0.841  &  0.821  & 6.27 x $10^6$   &   449.13 \\\hline
ADAM  &  RA  &  16  &  4  &  0.833  &  0.819  & 0.63 x $10^7$   &   111.20 \\\hline
ADAM  &  RA  &  8  &  1  &  0.848  &  0.821  & 3.14 x $10^6$   &   895.85 \\\hline
ADAM  &  RA  &  8  &  4  &  0.830  &  0.822  & 0.31 x $10^7$   &   224.12 \\\hline
\textbf{GD}  &  \textbf{HE}  &  \textbf{2796}  &  \textbf{1}  &  \textbf{0.822}  &  \textbf{0.823}  & \textbf{2.82 x} $\mathbf{10^9}$   &   \textbf{1} \\\hline
GD  &  HE  &  16  &  1  &  0.836  &  0.826  & 1.58 x $10^7$   &   178.66 \\\hline
GD  &  HE  &  16  &  4  &  0.837  &  0.828  & 1.58 x $10^7$   &   42.07 \\\hline
GD  &  HE  &  8  &  1  &  0.835  &  0.828  & 7.92 x $10^6$   &   355.37 \\\hline
GD  &  HE  &  8  &  4  &  0.839  &  0.821  & 0.79 x $10^7$   &   90.22 \\\hline
\textbf{GD}  &  \textbf{RA}  &  \textbf{2796}  &  \textbf{1}  &  \textbf{0.833}  &  \textbf{0.822}  & \textbf{2.82 x} $\mathbf{10^9}$   &   \textbf{1} \\\hline
GD  &  RA  &  16  &  1  &  0.84  &  0.822  & 6.49 x $10^6$   &   433.50 \\\hline
GD  &  RA  &  16  &  4  &  0.835  &  0.819  & 0.65 x $10^7$   &   111.94 \\\hline
GD  &  RA  &  8  &  1  &  0.843  &  0.824  & 3.26 x $10^6$   &   862.77 \\\hline
GD  &  RA  &  8  &  4  &  0.834  &  0.836  & 0.33 x $10^7$   &   207.59 \\\hline
\end{tabular}}
\end{table}

\end{document}